\documentclass[paper,11pt]{article}
\usepackage{cite}
\usepackage{amsfonts}
\usepackage{amsmath}
\usepackage{amssymb}
\usepackage{graphicx}
\usepackage{placeins}
\usepackage{epsfig,subfigure}
\usepackage{epsfig}
\usepackage{color,float}
\usepackage{lscape}
\usepackage{dcolumn}
\usepackage{bm}
\usepackage{placeins} 

\DeclareMathOperator{\sinc}{sinc}
\DeclareMathOperator{\meas}{meas}
\newcommand{\dif}{\mathop{}\!\mathrm{d}}

\begin{document}
\title{Spatio-temporal chaos and clustering induced by nonlocal information and vaccine hesitancy in the SIR epidemic model}
\author{Malay Banerjee$^1$, Samiran Ghosh$^1$,  Piero Manfredi$^2$, Alberto d'Onofrio$^{3,4,5}$\thanks{A d'Onofrio is no more associated to IPRI.}
 }
\date{\today }
\maketitle
{\center$^1$ Department of Mathematics and Statistics, Indian Institute of Technology Kanpur, Kanpur 208016, India, \\[2mm]}
{\center$^2$ Department of Economics, Pisa University, Pisa, Italy, \\[2mm]}
{\center$^3$ Department of Mathematics and Geosciences, University of Trieste, Trieste, Italy\\[2mm]}
{\center$^4$ Institute \textit{Camille Jordan}, Universit\'{e} \textit{Claude Bernard} Lyon 1, Villeurbanne, France. \\[2mm]}
{\center $^5$ Formerly at: International Prevention Research Institute. 106 Cours Lafayette, 69006 Lyon, France\\[2mm]  }
{\center Corresponding Authors: Malay Banerjee and Alberto d'Onofrio.\\ 
Emails: malayb@iitk.ac.in, alberto.donofrio@units.it  \\[2mm]  }

\begin{abstract}
Human behavior, and in particular vaccine hesitancy, is a critical factor for the control of  childhood infectious disease. Here we propose a spatio-temporal behavioral epidemiology model where the vaccine propensity depends on information that is non-local in space and in time. 
The properties of the proposed model are analysed under different hypotheses on the spatio-temporal kernels tuning the vaccination response of individuals. 
As a main result, we could numerically show that vaccine hesitancy induces the onset of many dynamic patterns of relevance for epidemiology. In particular we observed: behavior-modulated patterns and spatio-temporal chaos. This is the first known example of human behavior-induced spatio-temporal chaos in statistical physics of vaccination. Patterns and spatio-temporal chaos are difficult to deal with, from the Public Health viewpoint, hence showing that vaccine hesitancy can cause them could be of interest. \textbf{Additionally, we propose a new simple heuristic algorithm to estimate the Maximum Lyapunov Exponent.}
\end{abstract}


\section{Introduction}
\label{intro}
A key area of Statistical Physics of Vaccination \cite{spva} has focused on the parents' immunization decisions \cite{manfredi2013modeling,spva} concerning childhood vaccine preventable infectious diseases, such as measles, pertussis and varicella. This problem, as old as the smallpox vaccine \cite{manfredi2013modeling,spva}, has blown-up in  recent decades due to the increasing phenomenon of \emph{vaccine hesitancy} that has been included by WHO among the most serious threats to global health \cite{macdonald2015vaccine}. Indeed, the high degrees of herd immunity achieved in the Western world at the beginning of the 2000s - after decades of sustained vaccination at steadily high coverage - have brought the  \textit{perceived} risks from these infection to such negligible levels that in the above mentioned comparisons between  real benefits and  perceived costs (e,g., vaccine side effects) of immunization are causing phenomena of parents' escape from vaccination because they erroneously evaluate that costs of vaccination \textbf{exceed} its benefits (the phenomenon of  \emph{pseudo-rational avoidance of vaccination}  \cite{manfredi2013modeling,domasa}).  

Among the many facets of vaccine hesitancy a central one deals with the relationship between individual's decisions, collective coverage, and the available information on disease and immunization that is used by parents to elaborate their decision to vaccinate/not to vaccinate their children.
As most vaccine preventable infectious diseases are endemic i.e., they persist over time by showing recurrent temporal behaviors and travel over space \cite{mayanderson,hethcote2000mathematics}, the amount and type of related information \textbf{are} widely variable over time and space.  

Space is also central to the understanding of the spread and control of Infectious Diseases (IDs) because of the non-trivial impact of humans'  spatial mobility. Indeed, \textit{'knowledge of the spatial distribution and temporal trends of diseases is
an important prerequisite for the effective application of preventive and interventive
measures in order to reduce corresponding disease burdens'} \cite{kramer2010modern}. Many approaches are possible to describe human mobility among which the simplest is the use of models based on reaction--diffusion paradigm \cite{mayanderson,capasso1993mathematical,hethcote2000mathematics,murray2002intro,murray2002spatial,keeling2011modeling,martcheva2015introduction,malchow,anita2017reaction}.

In this article, we aim at improving our understanding of the interplay between information on infectious diseases and vaccine hesitancy in relation to childhood immunization, by adding an explicit spatial dimension in the propensity to vaccinate  in the well-known SIR model with vaccine hesitancy \textbf{introduced in} \cite{domasa}.

In particular, we want to investigate how time and space-modulated changes of perceived risks with respect to the infection and the vaccine side effects can impact on the spatio--temporal dynamics of endemic childhood infectious diseases.

A key point of non–spatial behavioral epidemiology models, is that immunization decisions are seldom based on information on the current prevalence of the infection only. Indeed, agents typically also take past
information into account. Further, information on serious vaccine adverse events (VAEs) typically \textbf{includes} data (and rumors) going far back into the past. Consequently, the resulting mathematical models should be temporally non--local \cite{domasa,domapo11,spva}.

Similarly, non-spatially structured models suffer the shortcoming that they can include only information on the global level, a crude abstraction. However, vaccination decisions seldom \textbf{depend} on purely local or purely global information only. Rather, we expect that agents base their decisions on information collected at appropriate non--local scales, attributing less and less importance to infection prevalence on far distant areas. For example, in relation to measles immunization, a major early study of behavioral epidemiology found that the key determinant of measles vaccine uptake in the US was the recent measles prevalence in the State families lived in  \cite{philipson1996private}, providing first substantive evidence of the non-locality (and of the non-globality either) of the (delayed) information used.

Therefore, the appropriate spatio--temporal models of infection dynamics will be doubly non--local: both spatially and temporally.

Consistently, \textbf{here}  we will consider a model with a doubly non--local behavioral response by adding space to the widely used SIR model with vaccine hesitancy proposed in \cite{domasa}. 

Namely, we plug the above-mentioned assumption on both human mobility and on space-dependent information within the simplest space-explicit formal setting: namely the basic reaction-diffusion PDE model.

Our main goal is to explore if and how the use of non-local spatio-temporal information to inform vaccination responses can generate  rich dynamics e.g., clusters or other complicated spatio--temporal patterns, and if yes, to characterize them. Indeed, there is evidence from ecological modeling that non-local interactions can alter the spatial pattern formation scenario. A main result is represented by the stationary Turing pattern formation in the spatio-temporal version of the Rosenzweig-MacArthur  model \cite{MB_VV_2017_nonlocal} resulting from non-local interactions. More in general, non-local interactions in intra- and inter-specific competition can induce a wide range of dynamic patterns namely modulated travelling wave, quasi-periodic oscillation and spatio-temporal chaos \cite{Swadesh_BMB,MB_VV_chaos,merchant2015selection,MB_Lai_Zhang2016}.

This work is organized as follows: in section 2 (and in the Appendix) we summarize some background information; in Section 3 we introduce our model; in Section 4 we model the spatio--temporally non--local information index, a key ingredient of our model; in Section 5 we study the onset of Turing instability in absence of temporal non--locality; in Section 6 we investigate the impact of temporal non--locality, showing qualitatively the onset of spatio-temporal chaos; in section 7 we demonstrate the spatio-temporal chaotic nature of the simulated dynamics. Concluding remarks end this work.

\section{Background on the SIR model with vaccine hesitancy} \label{nospace}
In this section we briefly illustrate the SIR model with vaccine hesitancy \cite{domasa} and its properties. The SIR model with mandatory vaccination is, instead, summarized in the Appendix A.\\
In \cite{domasa} the following SIR model with vaccine hesitancy model was proposed
\begin{align} 
\dot s&=\mu \left(1-p(m(t))\right) -\mu s - \beta(t)  i s, \label{tpbS}\\ 
\dot i&= \beta(t) i s - (\nu+\mu) i, \label{tpbI}
\end{align}
where $s(x,t)$ and $i(x,t)$ are the densities of, respectively, susceptible and infectious subjects, $\mu$ is the death and birth rate, $\beta(t)$ is the transmission rate, $\nu$ is the rate of recovery (i.e., of exit from the infectious state).\\
This model includes vaccine hesitancy by representing the vaccine uptake $p$ as a positive and increasing function of a phenomenological  \emph{information index} $m(t)$ \cite{domasa}. The information index is an auxiliary state variable summarizing the information on the present and past spread of the infection (and its sequelae) that is available to  parents. The underlying idea is that the collective vaccine coverage at time $t$, $p(t)$, is not anymore a constant but a dynamic variable, depending on the overall risks from the infection perceived by  vaccine decision makers, where these risks are summarised by index $m$. Note that $m(t)$ does not straightforwardly correspond to a simple physical phenomenon, since the information collection and distribution is extremely complex. Phenomenological models of complex phenomena are widely used in physics of complex systems \cite{cross2009pattern,gros2010complex,hoekstra2010simulating,ivancevic2008complex,pismen2006patterns}, especially in theoretical biophysics \cite{murray2002intro,murray2002spatial} and in sociophysics sciences \cite{helbing2012social,galam2016sociophysics,helbing2010quantitative,chakrabarti2006econophysics}, which are very rich of emergent phenomena, as well as in many branches of engineering. In particular, the vast majority of models of the spread and control of infectious diseases are phenomenological models \cite{keeling2011modeling,spva}. \\ 
Model (\ref{tpbS})--(\ref{tpbI}) must be complemented by a suitable model for $m(t)$. Assuming that $m$ depends on the information on the current and past prevalence of the disease, it follows that
$$ m(t) =\int_0^{+\infty}K(z)i(t-z)\dif z,$$
where the \emph{kernel} $K(z) \ge 0$ (said 'memory' kernel) \cite{domasa,spva} is assumed to obey
$$ \int_0^{+\infty}K(z)\dif z =1.$$
The function $K(z)$ represents the 'weight' that agents attribute to past information (hence the denomination of 'memory' kernel for $k(z)$). If $K(z) = \delta(z)$, where $\delta(z)$ is the Dirac Delta function, then agents only consider current information on the disease prevalence.\\
Interestingly, the system has a disease-free equilibrium $DFE=(1-p_0,0)$ whose stability properties are independent on $K(z)$. Namely:  i)if
$$ (1-p(0))\frac{\beta_m}{\mu+\nu}<1 \Rightarrow  p(0)>p_{cr}=1-(\mu+\nu)/\beta_m ,$$
then the DFE is Globally Asymptotically stable (GAS); ii) if $p(0)<p_{cr}$ then the DFE is unstable. \\
The above elimination condition implies that elimination can never be achieved if the \emph{resilient fraction} of parents - those vaccinating regardless of the trends of infection and related disease - remains persistently below the elimination threshold.\\
Moreover, still independently from the adopted kernel, if $\beta(t)$ is constant and $p(0)<p_{cr}$, then there is a unique endemic equilibrium $ EE =(s_e,i_e,m_e)$,  where $s_e = (\mu+\nu)/\beta $, $m_e=i_e$
and $i_e$ is the unique solution of the equation $ p(i)= p_{cr} -i $.\\
Unlike the DFE, the stability of the endemic equilibrium critically depends on the memory kernel $K(z)$. In \cite{domasa} two types of memory Kernels were used: a) the Dirac delta kernel centered at time $t$, that implies 
$ m(t)=i(t)$ i.e. vaccine decisions are taken by only using only current information on infection prevalence; b) the exponentially fading kernel (EFK)
$$ K(z)= a e^{-a z},$$
implying that the information taken into account to make vaccine decisions exponentially declines. In this case, the average memory time is $T=1/a$. As known from the basic theory of delay systems \cite{macdonald2013time}, the EFK allows finite dimensional reduction of the model since it holds: 
$$ \dot m = a (i-m).$$
If  $K(t)=\delta(t)$ then the endemic equilibrium is GAS \cite{domasa}, while in the case of EFK the system may exhibit sustained oscillations by a Hopf bifurcation of the endemic state, yielding to recurrent behaviour--induced epidemics \cite{domasa}. 

\section{A spatio--temporal model of voluntary vaccination and its equilibria}\label{mode}
The greatest limitation of all non-spatial epidemic models, including also (\ref{tpbS})-(\ref{tpbI}), is that they do not take into the account spatial heterogeneity. The first source of such heterogeneity is of course human mobility. Many modeling options could be adopted \cite{keeling2011modeling}: patch models, lattice-gas cellular automata, individual based models etc. Among those a simple yet effective way is to adopt the reaction-diffusion framework \cite{keeling2011modeling} which is also coherent with the  approach of non-spatial mathematical epidemiology where contagion is abstracted by means of the mass action law of chemical physics \cite{keeling2011modeling,spva}. As stressed by Keeling and Rohani: \textit{'such models are generally used to provide theoretical predictions and a
generic understanding of the spatial spread of infection'}. Operationally, Reaction-Diffusion approach \cite{murray2002intro,murray2002spatial} allows to straightforwardly generalize in the spatial setting a non-spatial epidemic or ecologic or demographic model $\dot P = F(P)$, where $P(t)$ is the populations size or fraction at time $t$, to \cite{murray2002intro,murray2002spatial} $\partial_t p = D \nabla^2 p + F(p)$, where $p(x,t)$ is the spatial density of the population, and $D$ is the diffusion coefficient of subjects belonging to the populations.\\ 
Building on top of \ref{tpbS})-(\ref{tpbI}), we propose here the following spatio-temporal model of vaccine hesitancy:
\begin{align}
\partial_t s&=D \nabla^2 s +\mu \left(1-p(m(x,t))\right) -\mu s - \beta(t)  i s ,\label{tpbspS}\\
\partial_t i&=D \nabla^2 i +\beta(t) i s - (\nu+\mu) i, \label{tpbspI}
\end{align}
Here $s(x,t)$ and $i(x,t)$ denote the densities of susceptible and infectious subjects, and human mobility is included by means of the diffusion terms $D \nabla^2 s$ and $D \nabla^2 i$.\\
Finally, vaccine uptake here depends on a space-structured information index $m(x,t)$. This \emph{space-time information index}, summarises current and past information available - over the entire space - to parents. Model (\ref{tpbspS})-(\ref{tpbspI}) must be complemented by a model:
$$ 
   m(x,t) = \Psi[i(.,.)],
$$
where $\Psi[i(.,.)]\ge 0$ is a functional involving both space and time, and such that i) $ \Psi[0]= 0$; ii) if $f(.,.)<h(.,.)$ then $\Psi[f(.,.)]<\Psi[h(.,.)]$; iii) if $i(x,t)$ is temporally and spatially constant $i_e$  (i.e. at a homogeneous equilibrium) then
$$ m(x,t)= g(i_e) = constant, $$
where $g(u)$ is a non-negative increasing function of $u\ge 0$, for example $g(u)=u$.\\
Interestingly, some general results are independent from the specific form of $\Psi[i(.,.)]$. Namely:  i) it exists a spatially homogeneous disease free solution
$ DFE(x,t) = (1-p(0),0)$, ii) if 
$$ 
	(1-p(0))\frac{\overline{\beta}}{\mu+\nu}>1
$$
then the disease free solution $ DFE(x,t)$ is unstable. This can be immediately shown by linearization of model (\ref{tpbspS})--(\ref{tpbspI}) at the $DFE$ ; iii) if 
\begin{equation}\label{tbpspGAS}
	(1-p_0)\frac{\overline{\beta}}{\mu+\nu}<1
\end{equation}
then $DFE(x,t)$ is GAS. This can be easily seen from the following differential inequality
$$ \partial_t s <D \nabla^2 s +\mu \left(1-p_0\right) -\mu s  ,$$
implying by comparison that asymptotically in time $ s(x,t)<1-p_0 $,  which, in turn, implies that asymptotically in time
$$ \partial_t i < D \nabla^2 i +i\left(\beta(t)(1-p(0))  - (\nu+\mu) \right), $$	
implying that $i(x,t)\rightarrow 0$ and in turn $s(x,t) \rightarrow 1-p_0$.\\
Finally, if the transmission rate is homogeneous both in space and time, it is easy to show that the model has a spatially uniform Endemic Equilibrium
$$ EE(x,t)= (s_e,i_e), $$
where $s_e$ and $i_e$ are the same obtained for the non--spatial behavioral SIR model. As we will see, for the study of the local stability of $ EE(x,t)$ and for the simulations of the model, the specific form of $\Psi$ is fundamental.

\section{Modelling the space-time information index $m(x,t)$}\label{modelM1}
As argued in the introduction, the non--locality affecting the vaccination decision-making process is double. On the one hand, such decisions are rarely based on purely local information. Actually, the space range of the adopted information can be large and, sometimes, can involve all the domain of interest, as it happens e.g., with nation--wide information. However, in most relevant cases the weight attributed by decision-makers to information from different spatial sites \textbf{is} by no means uniform. 
Moreover, in the previous sections we have discussed as actual decision-making will seldom be based on current information only: they most-often \textbf{take} into account an appropriate summary of past information.\\ 
To cope with this double non-locality issue, we consider here the following doubly non--local form for the space-time information index:
\begin{equation}\label{modelM} 
	m(x,t) =\int_{0}^{+\infty}Q_{time}(\tau)\left(\int_{y \in \Omega}Q_{space}(y)i(x-y,t-\tau)\dif y\right)\dif \tau,
\end{equation}
which should capture \textbf{the most-frequent} mental model of humans in handling information over space-time. 

Formula (\ref{modelM}) involves a pair of independent kernels $(Q_{time}(\tau), Q_{space}(y))$. These kernels will be the key ingredient of the specific models proposed in the subsequent sections.

As for $Q_{time}(\tau)$ and $Q_{space}(y)$ we assume the following: i) $ Q_{time}(\tau)$ has the same meaning and properties of the pure time kernel $K(z)$; ii) $Q_{space}(y) \ge 0 $ is such that $Q_{space}(0)>0$,
$$ \int_{y \in \Omega}Q_{space}(y)\dif y =1,$$
and it models the weight that agents attribute to local and non--local information on the infection prevalence.\\
Here and in the following sections we will assume that the transmission rate is constant $\beta(t)=\beta$. Thus, the model has \textbf{a unique} spatially homogeneous endemic equilibrium $EE=(s_e,i_e)$, which takes the same values of the endemic equilibrium of the non-spatial model and of which one has to study the local stability.\\
Let us linearize the generic model (\ref{tpbspS})--(\ref{tpbspI})--(\ref{modelM})
at EE $ (s,i) = EE + (u,v)$ and consider the case where the square root of the diffusion coefficient is much smaller than the characteristic spatial scale of $\Omega$. In such a case, denoting as $(\widehat{u},\widehat{u})$ the Fourier transform (see Appendix B) of $(u,v)$ and as $\widehat{Q}_{space}(\xi)$ the one of $Q_{space}(x)$, yields:
\begin{align}
	\lambda \widehat{u} &=-\left(D \xi^2 +\mu+\beta i_e \right) \widehat{u}  -\left(\mu+\nu + \mu p^{\prime}(i_e)\widehat{Q}_{space}(\xi)\widehat{Q}_{time}(\lambda)  \right) \widehat{v} ,\nonumber\\
	\lambda \widehat{v}&=\beta i_e \widehat{u} - D \xi^2 \widehat{v} , \nonumber
\end{align}
whose associated characteristic equation reads as follows:
\begin{equation}\label{dispersionGeneral}
	\lambda^2 + ( 2 D \xi^2 +\mu +\beta i_e )\lambda + D^2 \xi^4 + (\mu +\beta i_e)D \xi^2 + \beta i_e \left(\mu + \nu + \mu p^{\prime}(i_e) \widehat{Q}_{space}(\xi) \widehat{Q}_{time}(\lambda)\right) =0.
	\end{equation}
In case of non--small $D$, one has to apply the Fourier series decomposition and obtain a similar equation where the eigenvalues depend on the Fourier quantized vector.\\
In section \ref{space} we analyze the particular case where only spatial information is non-local, whereas in section \ref{time} we will move to the general case of full non-locality in both space and time.
\section{Onset of Turing instability in absence of temporal non--locality: analytical and numerical results}\label{space}	
\begin{figure}[thpb]
\begin{center}
		\mbox{\subfigure[]{\includegraphics[width=5.5cm,height=6cm]{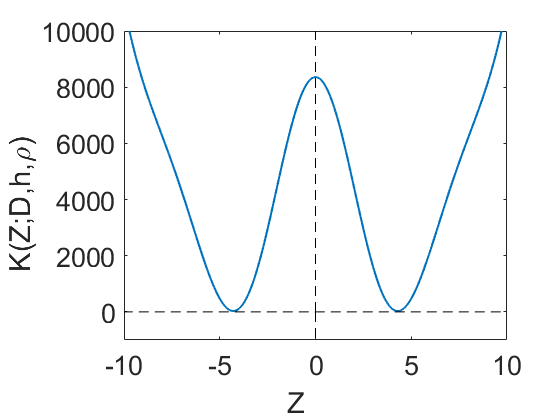}}
		\subfigure[]{\includegraphics[width=5.5cm,height=6cm]{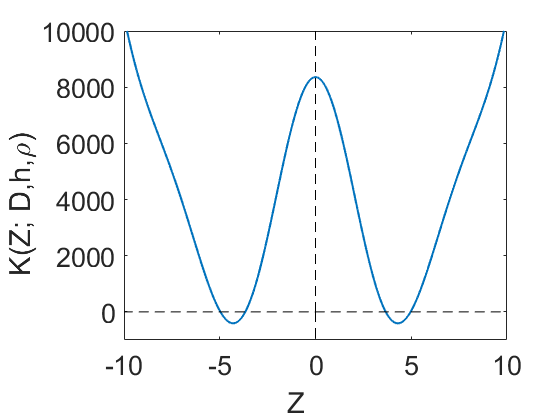}}
			\subfigure[]{\includegraphics[width=5.5cm,height=6cm]{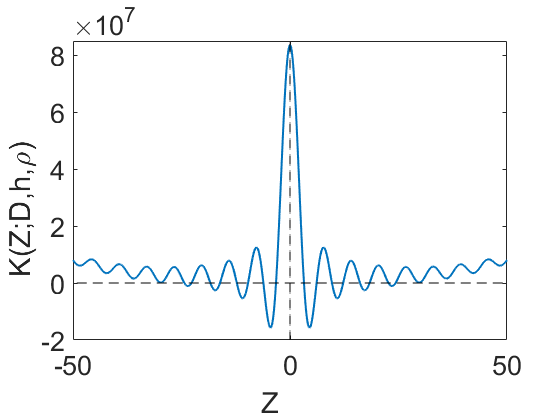}}
		}
\caption{Plot of the function $K(z; D, h,\rho)$ in (\ref{K}) are shown for $h=100$ and different choice of $D$  and $\rho$. The values of $D$,  $\rho$ are chosen in such a way that the minimum of the function $K(z; D, h=100,\rho)$ is negative, i.e., Turing instability emerges. In (a) $D=0.5$,  $\rho=6$, $i_e=0.000015$, in (b) $D=0.5$,  $\rho=9$, $i_e=0.000010$, and in (c) $D=0.005$,  $\rho=40$, $i_e=0.0000025$. The other parameter values are:\;
 $\mu =1/(75\times365)$, $\nu = 1/7$, $\beta=1.43$.}
\label{fig:plot_K}
\end{center}
\end{figure}

\begin{figure}[thpb]
\centerline{\includegraphics[width=12cm,height=6.5cm]{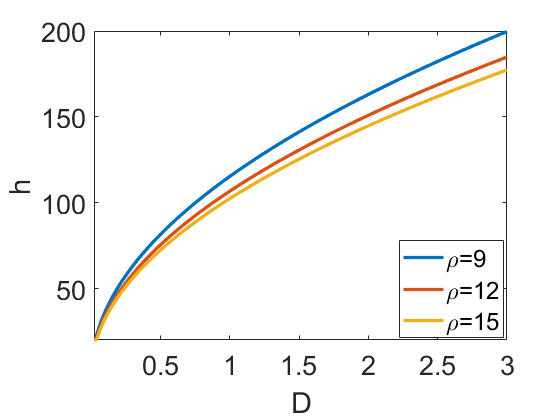}}
\caption{Bifurcation diagram in the $D-h$ parametric space. The upper region corresponds to the region of Turing instability and the lower region corresponds to the locally stable region for the endemic equilibrium. Other parameter values are:  $\mu =1/(75\times365)$, $\nu = 1/7$, $\beta=1.43$.}
\label{fig:instability_region}
\end{figure}

In this section, we will investigate the behavior of the system in the case where vaccine \textbf{decisions} are taken by only using the available spatially structured information on \emph{current} infection prevalence.
In such a scenario, it was  \textbf{shown} in \cite{domasa} that the endemic equilibrium of the non-spatially structured model is GAS.\\
In the spatio--temporal setting the dispersion equation  (\ref{dispersionGeneral}) becomes the following second-order algebraic equation:
\begin{equation}\label{dispersionOnlyQspace}
	\lambda^2 + ( 2 D \xi^2 +\mu +\beta i_e )\lambda +  b_0(\xi) =0,
	\end{equation}
where,
\begin{equation}\label{c0}
b_0(\xi)=D^2 \xi^4 + (\mu +\beta i_e)D \xi^2 + \beta i_e \left(\mu + \nu + \mu p^{\prime}(i_e) \widehat{Q}_{space}(\xi) \right).
	\end{equation}
It is important to remind that $i_e$ is a function of the model parameters.\\	
From the pair \eqref{dispersionOnlyQspace}-\eqref{c0} we note the following i) if the spatial kernel is positive on the whole set $\Omega$, then the spatially homogeneous Endemic Equilibrium EE remains LAS: this is the case of both the Dirac Delta kernel ( $Q_{space}(y)=\delta(y)$) and of the Gaussian kernel $(Q_{space}(y),\widehat{Q}_{space}(\xi))  = (A e^{-a y^2}, e^{-\xi^2/a})$; ii) if the spatial kernel is null outside a maximum area and it can assume negative values (see later for an example), so that if for some $\xi$ it holds that 
$$
b_0(\xi)<0,
$$
then the spatial symmetry is broken and a Turing pattern arises \cite{cross2009pattern,pismen2006patterns,murray2002spatial}. As expected from the non-spatial analysis in \cite{domasa}, a particularly steep vaccination response function $p(.)$ at the endemic equilibrium  favors the onset of spatial instability. \\
If $ \Omega$ is bounded and $Q_{space} = 1/\meas(\Omega)$ (where $\meas(\Omega)$ is the measure of $\Omega$)
$$  m(x,t)=\frac{1}{\meas(\Omega)}\int_{\Omega}i(x,t)\dif x,$$
i.e. $m(x,t)$ is the average value of the prevalence and the information has no spatial components: $\partial_x m(x,t)=0$, then $c_(\xi_n)=\delta_0$ for $n\neq 0$ and $c_0(0)=\delta_0 + Z $. This implies that local stability prevails for all modes.\\
Let us focus on the possibility of Turing-type instabilities. Let us first consider the case $D=0$, i.e., full absence of spatial movement. In such a case a Turing bifurcation occurs provided that:
$$ 
\mu + \nu + \mu p^{\prime}(i_e) \widehat{Q}_{space}(\xi) <0.
$$
Let us now consider the following piece--wise linear form for the vaccine uptake $p(m)$ \cite{domasa} 
$$
p(m)=p_0 +\min\left(  c m , 1-p_0 \right),    
$$
implying $ i_e=i_e(c)$. Assume further that the 'top--hat' kernel is used, which is defined as follows over $\mathbb{R}^n$, $n=1,2$
$$  Q_{space}(y)=B_n  Hev(h-|x|) $$
where $ B_1= 1/(2h)$ and $B_2= (1/(\pi h^2))$, and whose Fourier transform is as follows
$$ \widehat{Q}_{space}(\xi)=\sinc(h\xi)=\frac{\sin(h\xi)}{h\xi}. $$
The function $\sinc(w)$ has its absolute minimum at  $w_m\approx 4.5$, where $\sinc(w_m)\approx-0.21$, so that if
$$ \mu + \nu -0.21 \mu c <0, \Rightarrow c >  c_* \approx 4.761\left(1 +\frac{\nu}{\mu} \right) ,$$
then there is the onset of Turing instability.\\
Now, let us more in general consider,
$$ \mu + \nu + \mu c \sinc(h\xi) <0 \Rightarrow \sinc(h\xi) <- \frac{1}{c}\left(1 +\frac{\nu}{\mu} \right). $$
Note that for childhood infectious disease having a short infectious phase (and recalling that $1/\mu$, represents the average length of human life) the quantity $ 1+(\nu/\mu) $ has an order of magnitude greater than $10^3$. Setting $c=\rho(1+(\nu/\mu))$ with $\rho >0$, we may rewrite
$$
b_0(\xi)=D^2 \xi^4 + (\mu +\beta i_e)D \xi^2 + \beta i_e \left(\mu + \nu \right) \left(1 + \rho \sinc(h\xi) \right) .
$$
Defining $ z= h \xi$, we have,
$$ b_0(\xi)= \frac{D^2}{h^4} K(z; D, h,\rho), $$
where,
\begin{equation}\label{K}
 K(z; D, h,\rho)= z^4 + \frac{h^2}{D}(\mu +\beta i_e)z^2 + \frac{h^4}{D^2} \beta i_e \left(\mu + \nu \right) \left(1 + \rho \sinc(z) \right) .
\end{equation}
The plot of the function $K(z; D, h,\rho)$ is shown in Fig.~\ref{fig:plot_K} for values of $D$ and $\rho$ such that there are intervals where the function $K(z; D, h,\rho)$ is negative in some intervals, i.e. a Turing bifurcation occurs. Note that in \textbf{the left} panel the instability is concentrated in a very narrow range of frequencies, whereas in the right panel there are multiple intervals of frequencies where there Turing instability occurs. Moreover, for some particular choice of $\rho$ and $h$, the threshold value $D=D_c$ for the onset of Turing instability is shown in the Table.~\ref{table_K}. Using formula (\ref{K}), Fig.~\ref{fig:instability_region} shows the associated bifurcation diagram in the parametric space $D-h$.

\begin{table}[thpb]
\begin{center}
\caption {Onset of Turing instability for different choice of $D$, $h$, and $\rho$ (using the formula \ref{K})}
\begin{tabular}{|c|c|c|c|c|} \hline
$\rho$ & $h$ & $D_c$ for which & $z_c$ \big(where $\{ K(z;D_c,h,\rho)\}$ & $\xi_c=z_c/h$ \\
 &  &  min$\{ K(z;D_c,h,\rho)\}=0$ & attains minimum \big) &  \\
 \hline
     & $50$ & $0.1882$ & $ \pm 4.1222$ & $\pm 0.0824$ \\\cline{2-5}
$9$  & $100$ & $0.7526$ & $\pm 4.1224$ & $\pm 0.0412$ \\\cline{2-5}
    & $150$ & $1.6934$ & $\pm 4.1223$ & $\pm 0.0275$ \\\cline{1-5}
     & $50$ & $0.2198$ & $ \pm 4.0363$ & $\pm 0.0807$ \\\cline{2-5}
$12$  & $100$ & $0.8792$ & $\pm 4.0363$ & $\pm 0.0404$ \\\cline{2-5}
    & $150$ & $1.9784$ & $\pm 4.0362$ & $\pm 0.0269$ \\\cline{1-5}
      & $50$ & $0.2386$ & $ \pm 3.9861$ & $\pm 0.0797$ \\\cline{2-5}
$15$  & $100$ & $0.9544$ & $\pm 3.9861$ & $\pm 0.0399$ \\\cline{2-5}
    & $150$ & $2.1473$ & $\pm 3.9861$ & $\pm 0.0266$ \\\cline{1-5}
 
\end{tabular}
\label{table_K}
\end{center}
\end{table}

\begin{figure}[thpb]
\begin{center}
		\mbox{\subfigure[]{\includegraphics[scale=0.55]{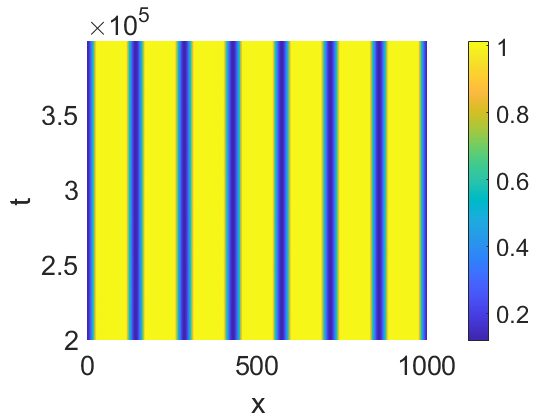}}\subfigure[]{\includegraphics[scale=0.55]{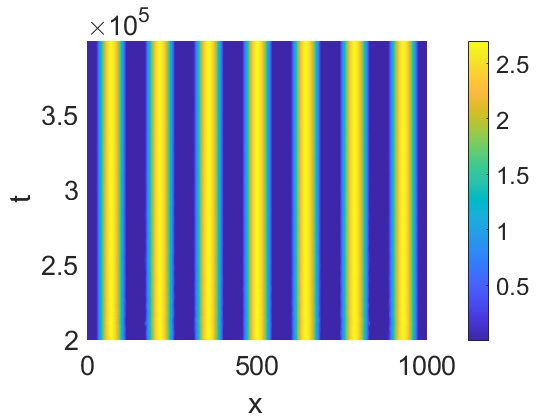}}}
\caption{Turing pattern in one dimension in the case $D=0$ Panel (a) shows  $s(x,t)/s_e$ and panel (b)  $i(x,t)/i_e$. This simulation is performed for $p(M)=p_0 +\textrm{min}\left(  c M , 1-p_0 \right)$. Parameter values: \;
$D=0$, $\mu =1/(75\times365)$, $\nu = 1/7$, $\beta=1.43$, $\rho = 7$, $p_0=0.5$, $s_e=0.1$, $i_e=0.000013$, $h=100$ and $L=1000$. The initial perturbation is: $s(x,0)=s_e+0.00001$, $i(x,0)=i_e+0.00001$ for $x \in [485, 515]$ and $s(x,0)=s_e$, $i(x,0)=i_e$ elsewhere.}
\label{fig:pattern_s_i_D=0}
\end{center}
\end{figure}

\begin{figure}[thpb]
\begin{center}
		\mbox{\subfigure[]{\includegraphics[scale=0.55]{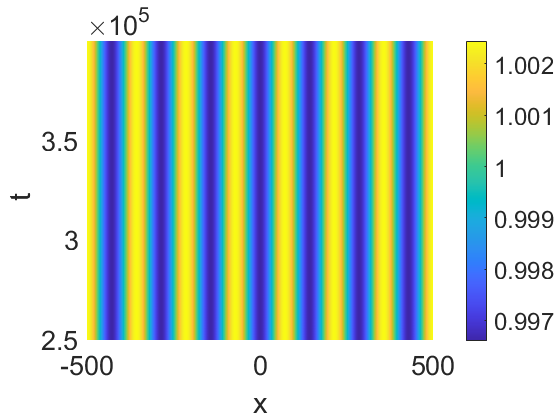}}
		\subfigure[]{\includegraphics[scale=0.55]{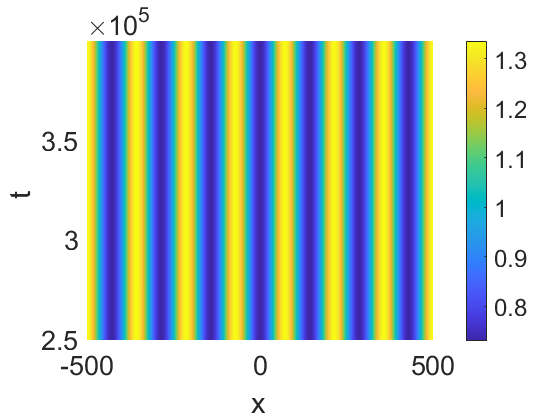}}}
\caption{Turing pattern in one dimension in the case $D=0.7>0$ Panel  shows $s(x,t)/s_e$ and panel (b) shows $i(x,t)/i_e$. This simulation is performed for $p(M)=p_0 +\textrm{min}\left(  c M , 1-p_0 \right)$. Parameter values:\;
$L=1000$, $D=0.7$, $\mu =1/(75\times365)$, $\nu = 1/7$, $\beta=1.43$, $\rho = 9$, $p_0=0.5$, $s_e=0.1$, $i_e=0.00001$, $h=100$. The initial perturbation is: $s(x,0)=s_e+0.001$, $i(x,0)=i_e+0.001$ for $x \in [-20, 20]$ and $s(x,0)=s_e$, $i(x,0)=i_e$ elsewhere.}
\label{fig:pattern_s_i_D=0.7}
\end{center}
\end{figure}

\begin{figure}[thpb]
\begin{center}
		\mbox{\subfigure[]{\includegraphics[scale=0.55]{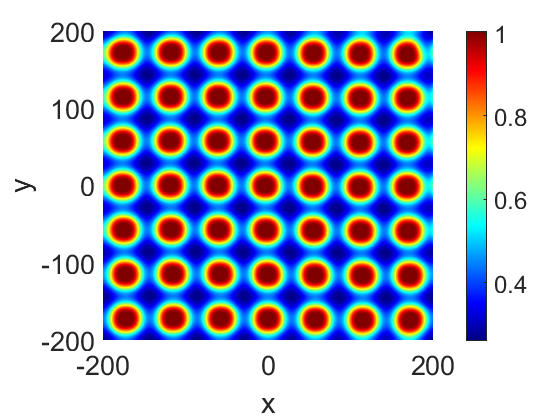}}
		\subfigure[]{\includegraphics[scale=0.55]{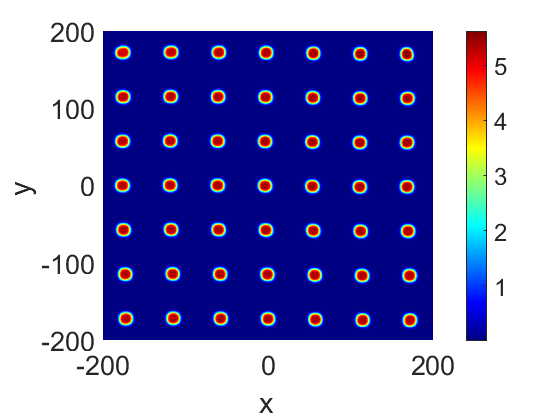}}}
\caption{Turing pattern in two dimensions in the case $D=0.7>0$ Panel  shows $s(x,t)/s_e$ and panel (b) shows $i(x,t)/i_e$. This simulation is performed for $p(M)=p_0 +\min\left(  c*M , 1-p_0 \right)$. Parameter values:\;
$L=400$, $D=0.005$, $\mu =1/(75\times365)$, $\nu = 1/7$, $\beta=1.43$, $\rho = 15$, $p_0=0.5$, $s_e=0.1$, $i_e=0.0000064$, $h=40$. The initial condition is: $s(x,y,0)=s_e+0.01\xi_{xy}^1$, $i(x,0)=i_e+0.00001\xi_{xy}^2$ where $\xi_{xy}^j$, ($j=1,2$) are two spatially uncorrelated white noise terms. Periodic boundary conditions are used. The patterns are obtained at $t=80,000$.}
\label{fig:pattern_2D_1}
\end{center}
\end{figure}

Now, we present the numerical simulation for the above illustrated scenario. To avoid boundary effects due to the non--local kernels, we performed all the simulations by assuming periodic boundary conditions. As per the vaccine hesitancy, We assumed that  $p(M)=p_0 +\min\left(  c M , 1-p_0 \right)$. Simulations illustrated in Fig.~\ref{fig:pattern_s_i_D=0} correspond to the case $D=0$ with \textbf{parameter} values that satisfies the condition of Turing instability. As predicted in the analytical results, stationary Turing patterns were obtained. The more significant case of $D \neq 0$ is shown in Fig.~\ref{fig:pattern_s_i_D=0.7} and Fig.~\ref{fig:pattern_2D_1} for, respectively, dimensions one and two. The stationary pattern observed for dimension one corresponds to a 'hot spot' pattern in two dimensions.

From the Public Health viewpoint, we can say that when the information used by parents is local w.r.t. time but nonlocal w.r.t. space may induce the emergence of strong spatial clusters of the disease.

\section{Impact of temporal non--locality: spatio-temporal chaos and static patterns}\label{time}
\begin{figure}[thpb]
\centerline{\includegraphics[width=8.25cm,height=5.25cm]{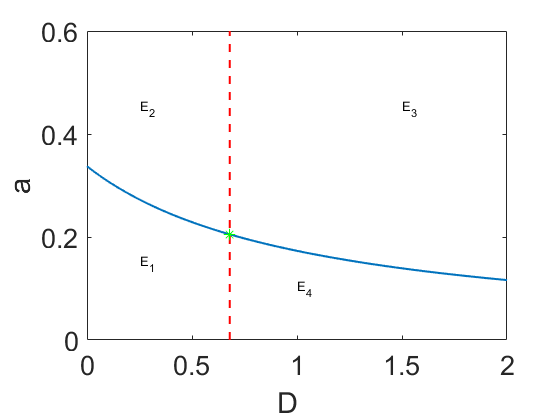}}
\caption{The bifurcation diagram $D-a$ parametric space, divided \textbf{into} 4 regions. Region  $E_1: \;$ Turing-Hopf Instability region, Region $E_2: \;$ Turing Instability region, Region $E_3: \;$ Stable region, Region $E_4:\;$ temporal Hopf region. The green point corresponds to the Turing-Hopf threshold $(D_c, a_c) =(0.68, 0.2051).$ Other parameter values are:  $\mu =1/(75\times365)$, $\nu = 1/7$, $\beta=1.43$ $h=100$, $\rho=8$ $p_0=0.5$. }
\label{fig:Turing_Hopf_region_1}
\end{figure}

Here, we will investigate the impact of vaccination decisions based  also on the past (spatially structured) information on the infection prevalence. We will focus on the already mentioned exponentially fading memory kernel $Q_{time}(t)= a e^{-a t}.$\\
In the purely temporal setting, temporal non--locality can destabilize the endemic equilibrium of the SIR model with vaccination decisions and trigger limit cycles via Hopf bifurcations \cite{domasa} \textbf{but chaos is not observed} \cite{domasa}. As a consequence, a number of scenarios can be obtained in our spatio-temporal model. In particular:
\begin{itemize}
    \item If in the purely temporal setting the endemic Equilibrium $EE$ is LAS, then in the spatio--temporal setting a Turing Instability (TI) may occur;
    \item If in the temporal setting $EE$ is unstable and limit cycles appear (but not temporal chaos: the transmission rate is constant \textbf{and the vaccine hesitancy does not induces chaos} \cite{domasa}) then in the spatio--temporal setting spatio--temporal chaos may appear.
\end{itemize}
\noindent
By applying the linear chain trick \cite{kuang} to the spatially- structured system, the model under study reads:
\begin{align}
\partial_t s&=D \nabla^2 s +\mu \left(1-p(m(x,t))\right) -\mu s - \beta  i s ,\label{tpbspSefk}\\
\partial_t i&=D \nabla^2 i +\beta(t) i s - (\nu+\mu) i, \label{tpbspIefk}\\
\partial_t m &= a \left(\int_{y \in \Omega}Q_{space}(y)i(x-y,t)\dif y - m\right).\label{tpbspMefk}
\end{align}
Since exponentially fading kernel has the following Laplace transform $ \widehat{Q}_{time}(\lambda) =a/(\lambda+a) $, then dispersion equation reads as follows:
$$
\lambda^3 + c_2(\xi) \lambda^2 + c_1(\xi) \lambda + c_0(\xi)=0,
$$
where,
$$ c_2(\xi)= a+ 2 D \xi^2 +\mu +\beta i_e >0,$$
$$ c_1(\xi)= \left( 2 D \xi^2 +\mu +\beta i_e \right)a + D^2 \xi^4 + (\mu +\beta i_e)D \xi^2 + \beta i_e (\mu + \nu)  >0, $$
$$ c_0(\xi)= a \left( D^2 \xi^4 + (\mu +\beta i_e)D \xi^2 + \beta i_e (\mu + \nu)  +  \beta i_e \mu p^{\prime}(i_e) \widehat{Q}_{space}(\xi)  \right).  $$
The Routh--Hurwitz conditions \textbf{give} that: i)Also here, if for some $\xi$ 
$$ c_0(\xi)<0$$
holds, then the endemic equilibrium is unstable and Turing pattern arises; ii) if for some $\xi$ it holds that 
$$ c_2(\xi) c_1(\xi)-c_0(\xi)<0,$$ i.e. if $a$ is such that:
$$
\left( D \xi^2 +\delta_1 \right)a^2 + \left( \left(D \xi^2 +\delta_1 \right)^2 - \delta_0 - Z \widehat{Q}_{space}(\xi)  \right)a +\delta_0 \left( D \xi^2 +\delta_1 \right) <0
$$
then Hopf instability occurs. The condition for Hopf instability can be written as follows
$$
\left( D \xi^2 +\delta_1 \right)^2 - \delta_0 - Z \widehat{Q}_{space}(\xi) <-2\sqrt{\delta_0}\left( D \xi^2 +\delta_1 \right)
$$
i.e.
$$
Z \widehat{Q}_{space}(\xi)>\left( D \xi^2 +\delta_1 \right)^2  + 2\sqrt{\delta_0}\left( D \xi^2 +\delta_1 \right)- \delta_0 . 
$$
The bifurcation diagram \ref{fig:Turing_Hopf_region_1}, shows that the  $D-a$ parametric space plane is divided in four regions: stability region, temporal Hopf region, Turing Instability region and Turing-Hopf instability region.

\begin{figure}[thpb]
\begin{center}
		\mbox{\subfigure[]{\includegraphics[scale=0.55]{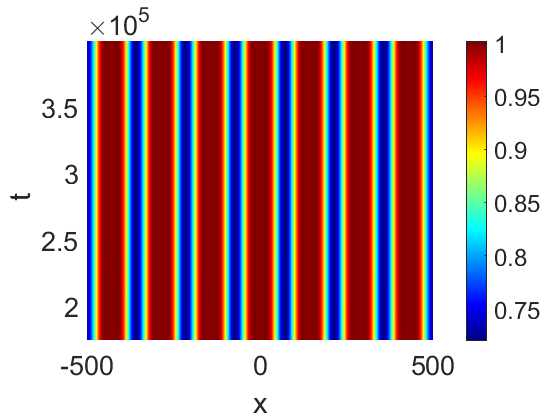}}
		\subfigure[]{\includegraphics[scale=0.55]{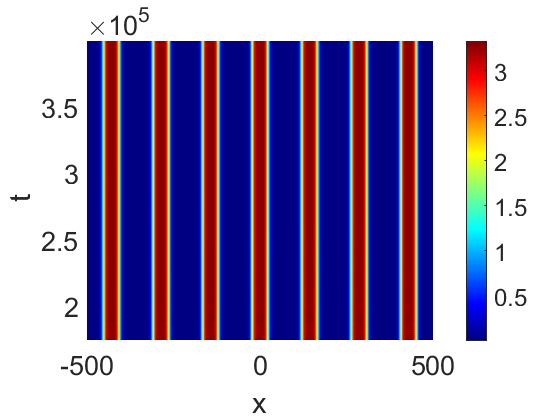}}}
\caption{ Impact of parameter $a$ on the dynamic behavior, in the one dimensional case. For $a=0.6$ a stationary Turing Pattern is observed. Panel  (a) shows  $s(x,t)/s_e$ and panel (b) shows $i(x,t)/i_e$. Parameter values:\;
$L=1000$, $D=0.05$, $\mu =1/(75\times365)$, $\nu = 1/7$, $\beta=1.43$, $\rho = 8$, $p_0=0.5$, $s_e=0.1$, $i_e=0.000011$, $h=100$. The initial perturbation is: $s(x,0)=s_e+0.0001\xi_x^1$, $i(x,0)=i_e+0.000001\xi_x^2$ for all $x$, where $\xi_x^1$ and $\xi_x^1$ are two spatially uncorrelated white noise terms. Periodic boundary conditions are used.}
\label{fig:exponentially_fading_1D_pattern}
\end{center}
\end{figure}

\begin{figure}[thpb]
\begin{center}
		\mbox{\subfigure[]{\includegraphics[scale=0.4]{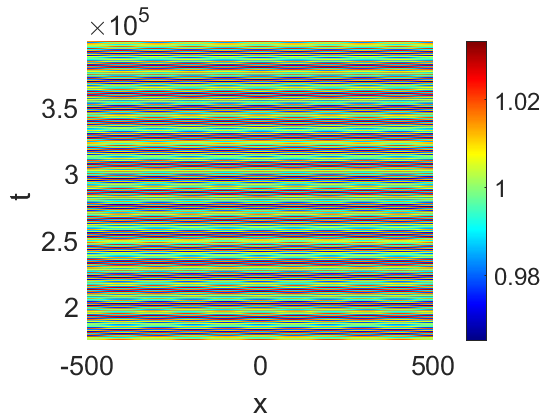}}
		\subfigure[]{\includegraphics[scale=0.4]{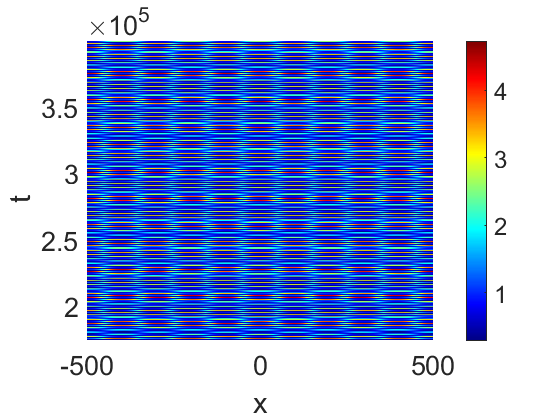}}
		\subfigure[]{\includegraphics[scale=0.4]{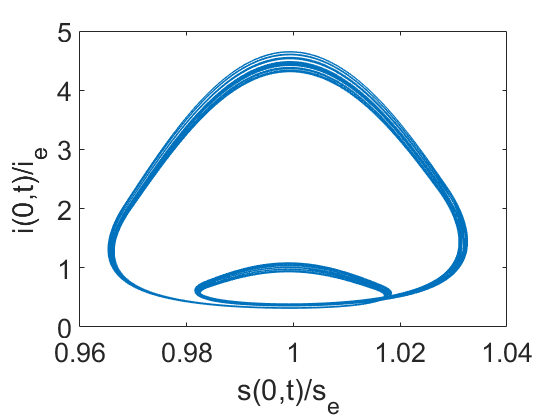}}}
		\mbox{\subfigure[]{\includegraphics[scale=0.4]{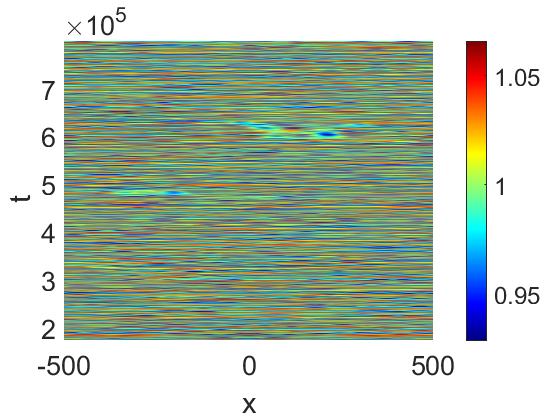}}
		\subfigure[]{\includegraphics[scale=0.4]{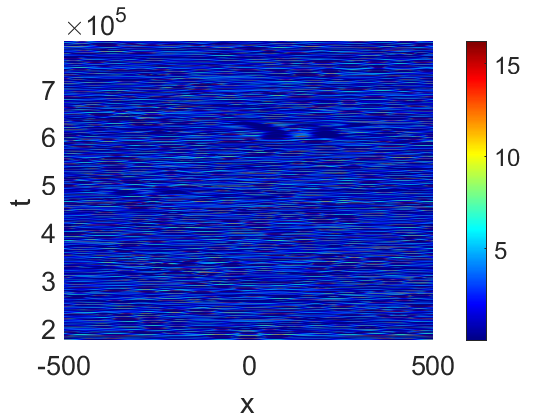}}
		\subfigure[]{\includegraphics[scale=0.4]{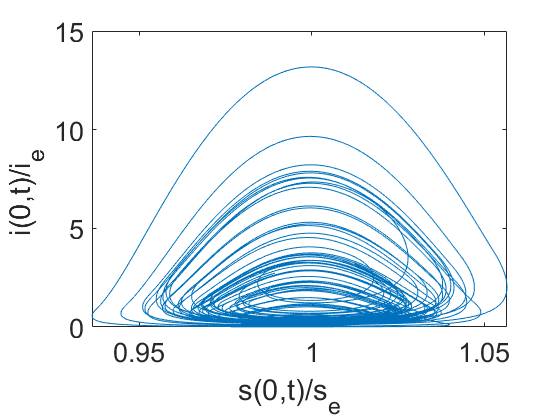}}}
		\mbox{\subfigure[]{\includegraphics[scale=0.4]{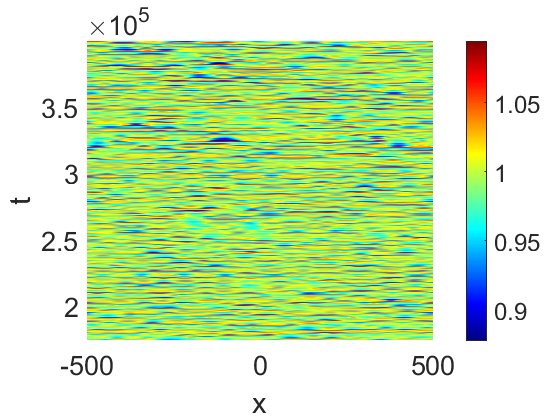}}
		\subfigure[]{\includegraphics[scale=0.4]{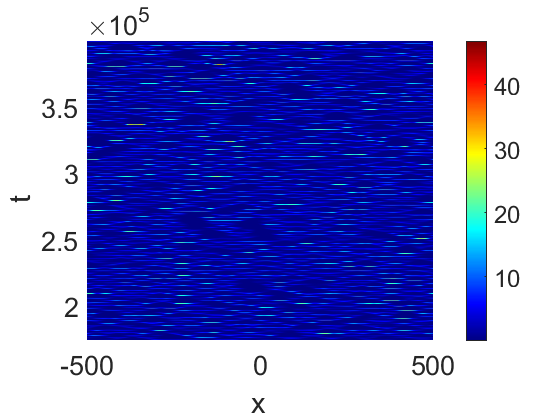}}
		\subfigure[]{\includegraphics[scale=0.4]{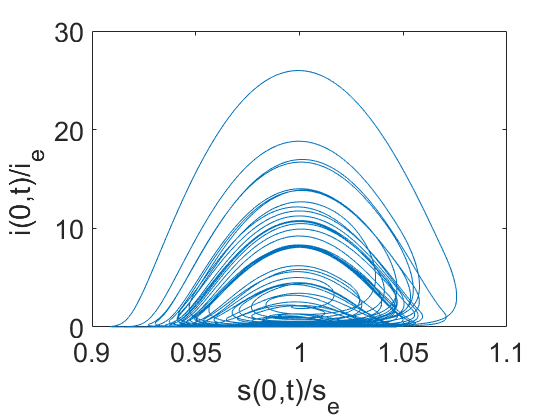}}
		}
\caption{ Transition to chaos: impact of parameter $a$ on the dynamic behavior, in the one dimensional case. From the first to the third rows the values of $a$ are as follows: $a=0.1, 0.033, 0.01$ Left panels show $s(x,t)/s_e$, central panels show $i(x,t)/i_e$, and right panels show the trajectories $( s(0,t)/s_e, i(t,0)/i_e)$.
$L=1000$, $D=0.05$, $\mu =1/(75\times365)$, $\nu = 1/7$, $\beta=1.43$, $\rho = 8$, $p_0=0.5$, $s_e=0.1$, $i_e=0.000011$, $h=100$. The initial perturbation is: $s(x,0)=s_e+0.0001\xi_x^1$, $i(x,0)=i_e+0.000001\xi_x^2$ for all $x$, where $\xi_x^1$ and $\xi_x^1$ are two spatially uncorrelated white noise terms. Periodic boundary conditions are used.}
\label{New1D}
\end{center}
\end{figure}

\begin{figure}[thpb]
\begin{center}
		\mbox{\subfigure[]{\includegraphics[scale=0.55]{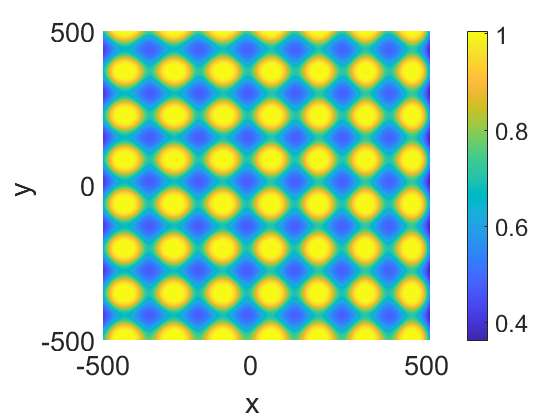}}
		\subfigure[]{\includegraphics[scale=0.55]{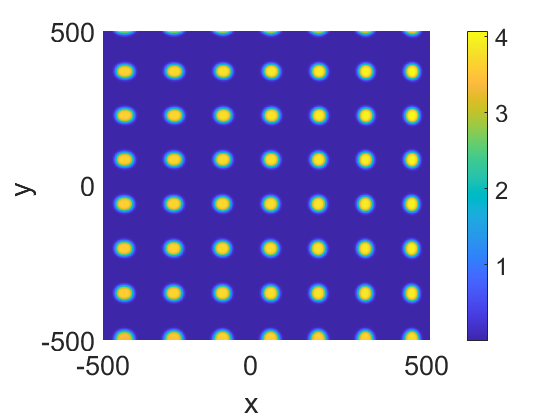}}}
\caption{Impact of parameter $a$ on the dynamic behavior, in the bi--dimensional case. For $a=0.6$ a stationary Turing Pattern is observed. Panel  (a) shows  $s(x,t)/s_e$ and panel (b) shows $i(x,t)/i_e$.  Parameter values:\;
$L=1000$, $D=0.05$, $\mu =1/(75\times365)$, $\nu = 1/7$, $\beta=1.43$, $\rho = 8$, $p_0=0.5$, $s_e=0.1$, $i_e=0.000011$, $h=100$ and $a=0.6$. The initial perturbation is: $s(x,y,0)=s_e+0.01\xi_{xy}^1$, $i(x,y,0)=i_e+0.00001\xi_{xy}^2$ where $\xi_{xy}^j$, ($j=1,2$) are two spatially uncorrelated white noise terms. Periodic boundary conditions are used. This pattern is obtained for $t=110000$.}
\label{Turing2Da0p6}
\end{center}
\end{figure}

\begin{figure}[htp]
\begin{center}
		\mbox{\subfigure[]{\includegraphics[scale=0.35]{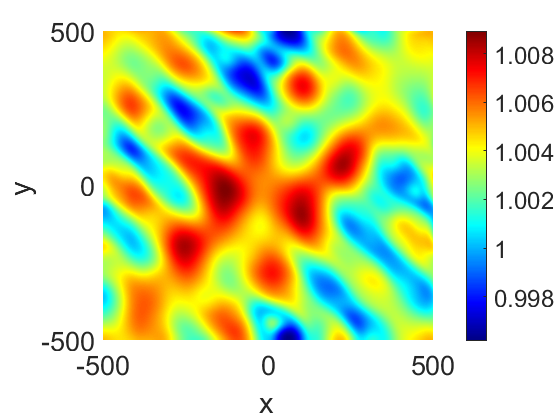}}
		\subfigure[]{\includegraphics[scale=0.35]{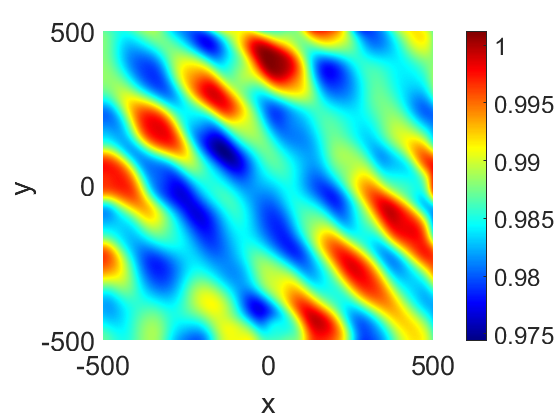}}
		\subfigure[]{\includegraphics[scale=0.35]{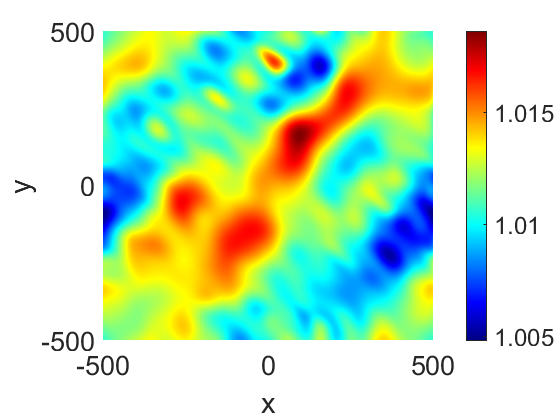}}}
		\mbox{\subfigure[]{\includegraphics[scale=0.35]{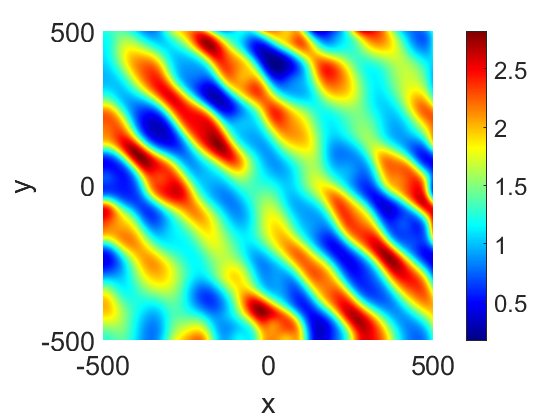}}
		\subfigure[]{\includegraphics[scale=0.35]{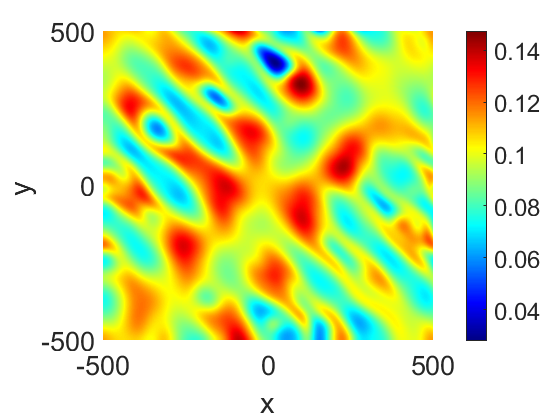}}
		\subfigure[]{\includegraphics[scale=0.35]{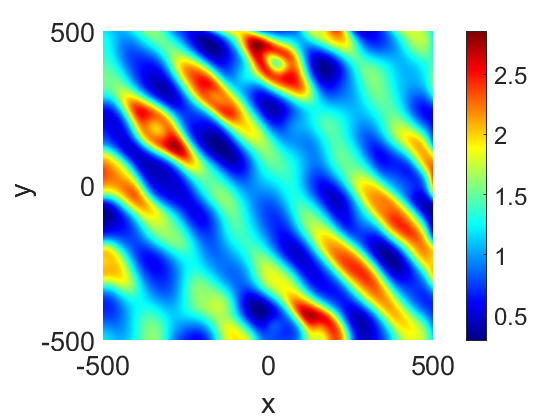}}}
			\mbox{\subfigure[]{\includegraphics[scale=0.35]{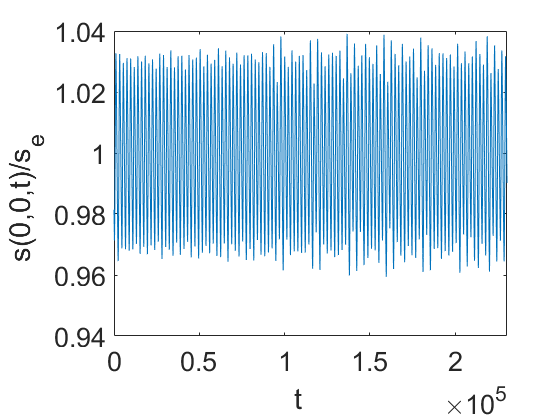}}
		\subfigure[]{\includegraphics[scale=0.35]{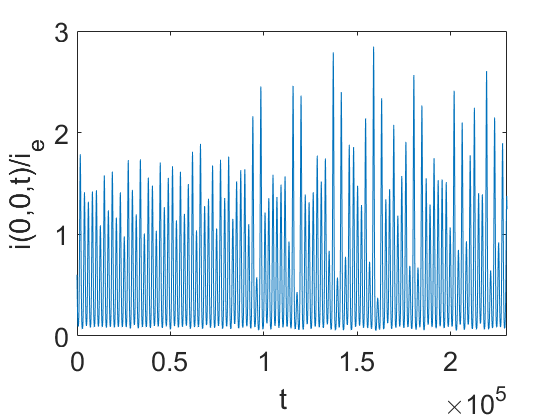}}
		\subfigure[]{\includegraphics[scale=0.35]{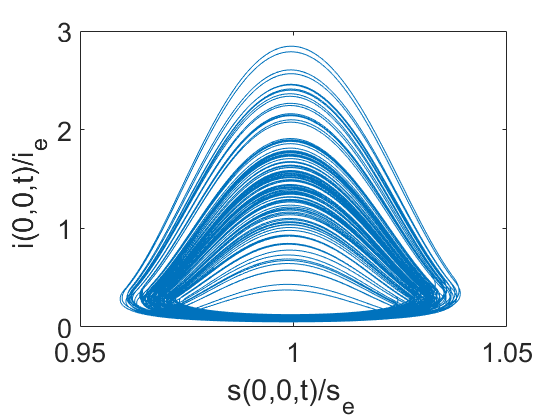}}}
		
			\mbox{\subfigure[]{\includegraphics[scale=0.35]{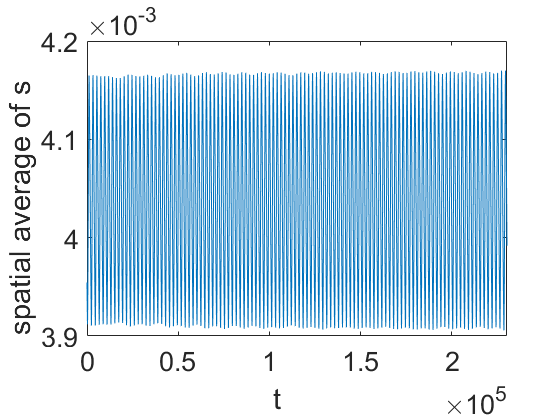}}
		\subfigure[]{\includegraphics[scale=0.35]{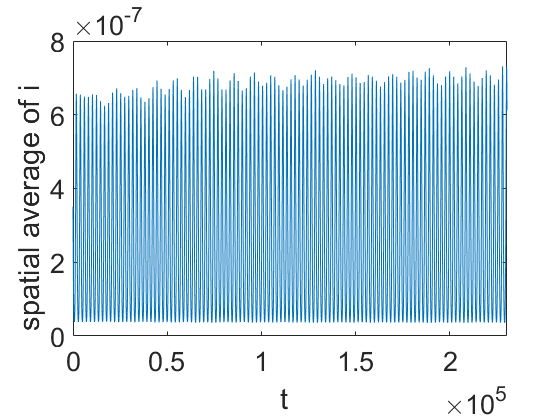}}
		\subfigure[]{\includegraphics[scale=0.35]{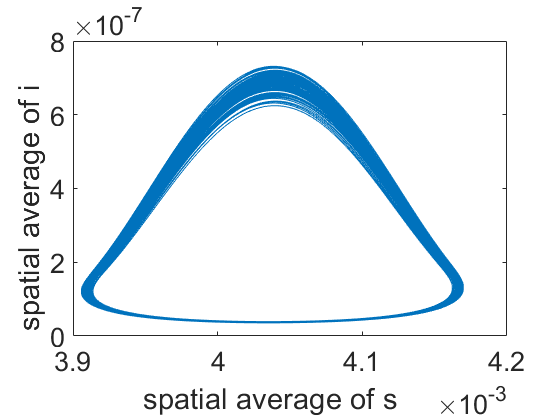}}}
\caption{Transition to spatio-temporal chaos. Impact of parameter $a$ on the dynamic behavior, in the bidimensional case. Here $a=0.1$. First and second rows show snapshots at three different times of the densities of, respectively, susceptible and infectious subjects, normalized to their respective endemic values. Third row shows time series and phase portraits computes in the spatial point $(0,0)$. Fourth row shows time series and phase of the averages state variables. Namely: panels (a), (d) correspond to snapshots of normalized value of $s(x,t)/s_e$  and $i(x,t)/i_e$ captured at time $t=250000$; panels (b), (e) show the same snapshots but  captured at time $t=252000$; finally panels (c), (f)  shows the snapshots captured at time  $t=254000$. Parameter values:\;
$L=1000$, $D=0.05$, $\mu =1/(75\times365)$, $\nu = 1/7$, $\beta=1.43$, $\rho = 8$, $p_0=0.5$, $s_e=0.1$, $i_e=0.000011$, $h=100$ and $a=0.1$. The initial perturbation is: $s(x,y,0)=s_e+0.01\xi_{xy}^1$, $i(x,y,0)=i_e+0.00001\xi_{xy}^2$ where $\xi_{xy}^j$, ($j=1,2$) are two spatially uncorrelated white noise terms. Periodic boundary conditions are used.}
\label{chaos2Da0p1}
\end{center}
\end{figure}

\begin{figure}[htp]
\begin{center}
		\mbox{\subfigure[]{\includegraphics[scale=0.35]{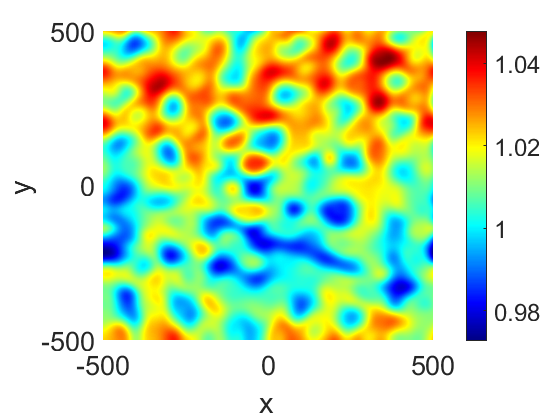}}
		\subfigure[]{\includegraphics[scale=0.35]{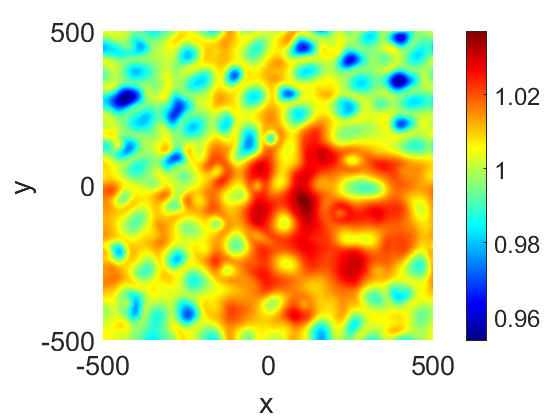}}
		\subfigure[]{\includegraphics[scale=0.35]{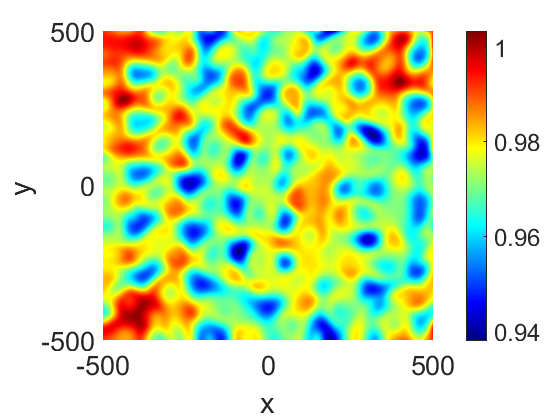}}}
		\mbox{\subfigure[]{\includegraphics[scale=0.35]{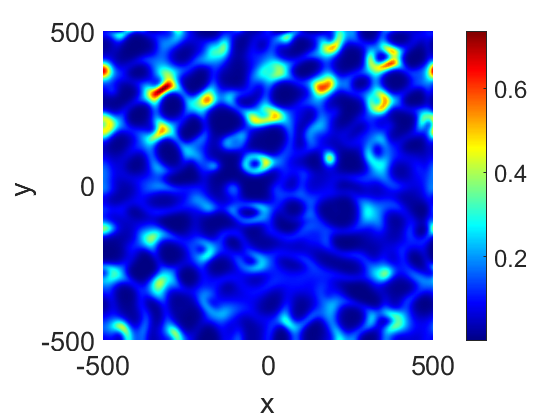}}
		\subfigure[]{\includegraphics[scale=0.35]{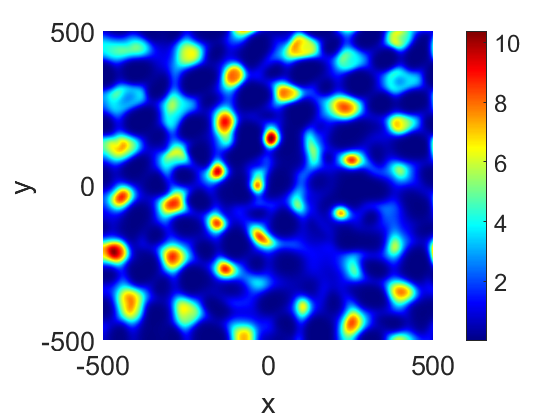}}
		\subfigure[]{\includegraphics[scale=0.35]{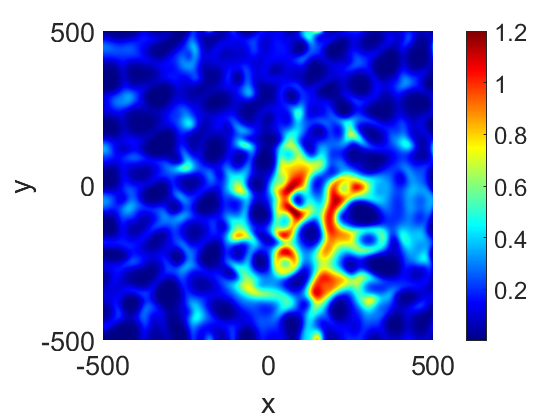}}}
			\mbox{\subfigure[]{\includegraphics[scale=0.35]{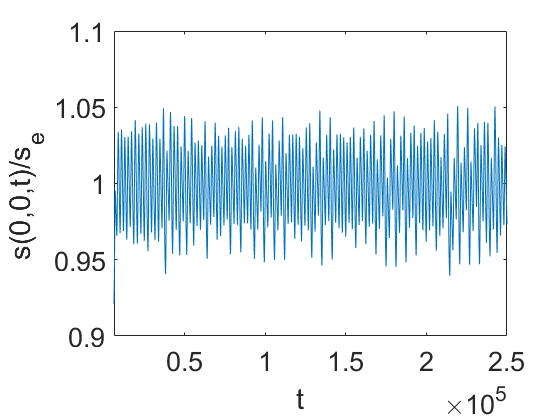}}
		\subfigure[]{\includegraphics[scale=0.35]{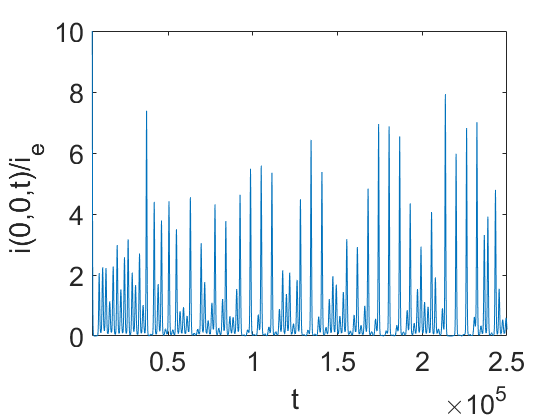}}
		\subfigure[]{\includegraphics[scale=0.35]{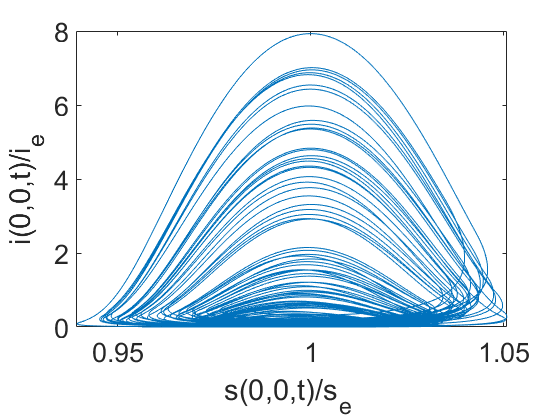}}}
		
			\mbox{\subfigure[]{\includegraphics[scale=0.35]{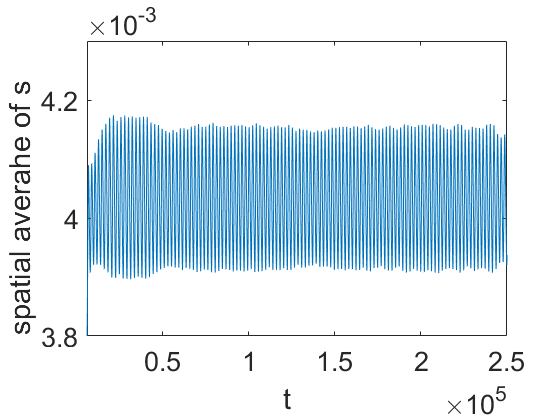}}
		\subfigure[]{\includegraphics[scale=0.35]{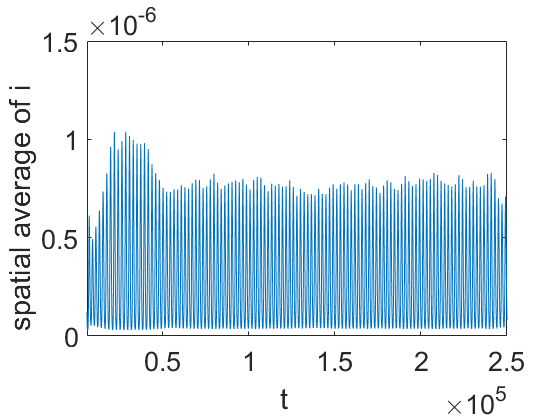}}
		\subfigure[]{\includegraphics[scale=0.35]{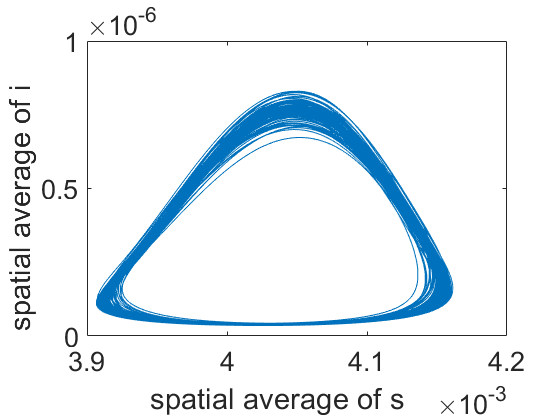}}}
\caption{Transition to spatio-temporal chaos. Impact of parameter $a$ on the dynamic behavior, in the bidimensional case. Here $a=0.033$. First and second rows show snapshots at three different times of the densities of, respectively, susceptible and infectious subjects, normalized to their respective endemic values. Third row shows time series and phase portraits computes in the spatial point $(0,0)$. Fourth row shows time series and phase of the averages state variables. Namely: panels (a), (d) correspond to snapshots of normalized value of $s(x,t)/s_e$  and $i(x,t)/i_e$ captured at time $t=250000$; panels (b), (e) show the same snapshots but  captured at time $t=252000$; finally panels (c), (f)  shows the snapshots captured at time  $t=254000$. Parameter values:\;
$L=1000$, $D=0.05$, $\mu =1/(75\times365)$, $\nu = 1/7$, $\beta=1.43$, $\rho = 8$, $p_0=0.5$, $s_e=0.1$, $i_e=0.000011$, $h=100$ and $a=0.033$. The initial perturbation is: $s(x,y,0)=s_e+0.01\xi_{xy}^1$, $i(x,y,0)=i_e+0.00001\xi_{xy}^2$ where $\xi_{xy}^j$, ($j=1,2$) are two spatially uncorrelated white noise terms. Periodic boundary conditions are used.}
\label{chaos2Da0p033}
\end{center}
\end{figure}

\begin{figure}[htp]
\begin{center}
		\mbox{\subfigure[]{\includegraphics[scale=0.35]{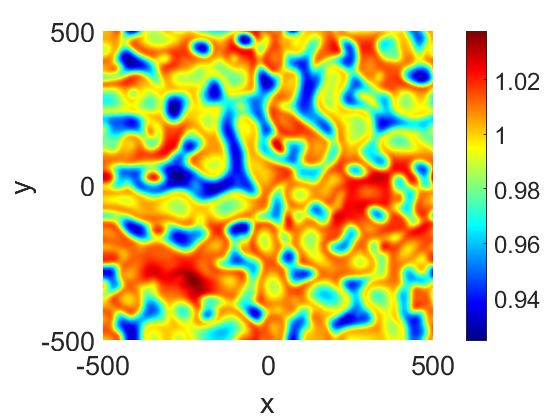}}
		\subfigure[]{\includegraphics[scale=0.35]{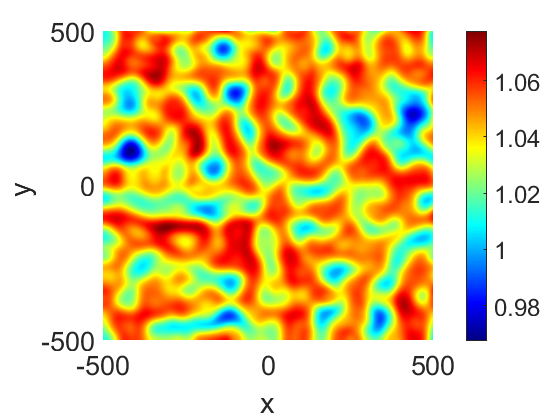}}
		\subfigure[]{\includegraphics[scale=0.35]{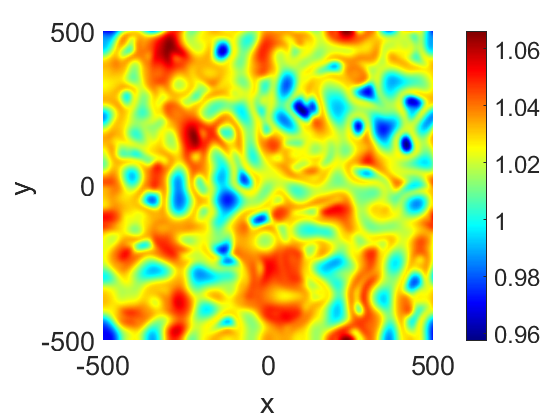}}}
		\mbox{\subfigure[]{\includegraphics[scale=0.35]{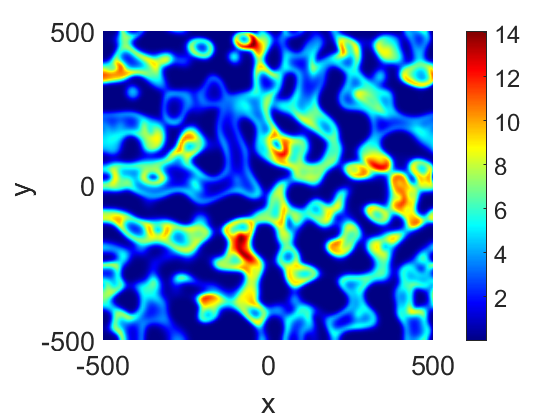}}
		\subfigure[]{\includegraphics[scale=0.35]{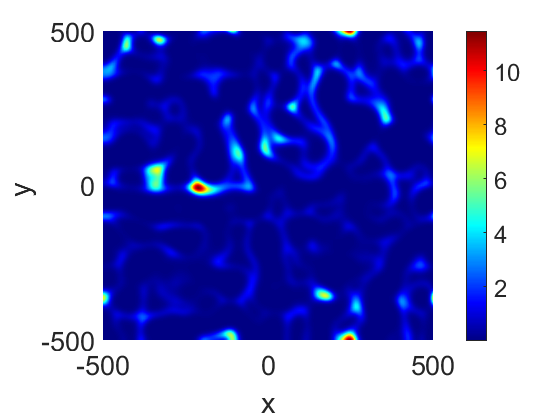}}
		\subfigure[]{\includegraphics[scale=0.35]{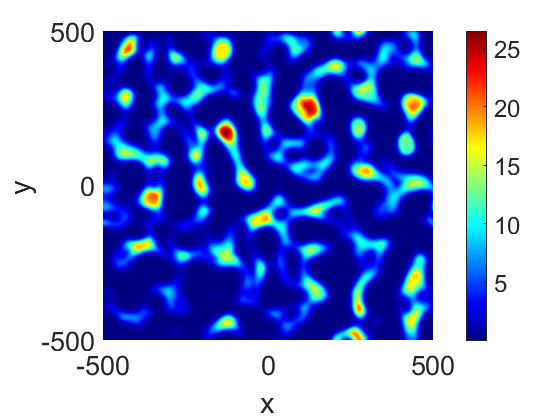}}}
			\mbox{\subfigure[]{\includegraphics[scale=0.35]{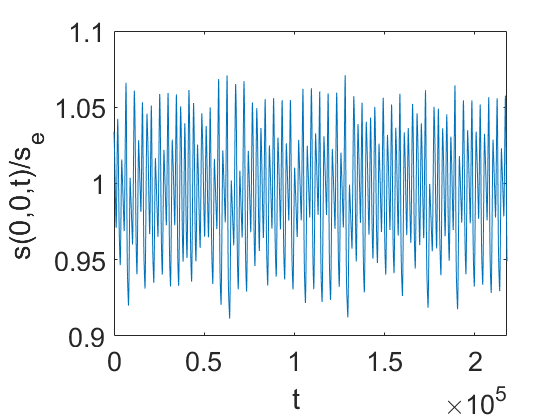}}
		\subfigure[]{\includegraphics[scale=0.35]{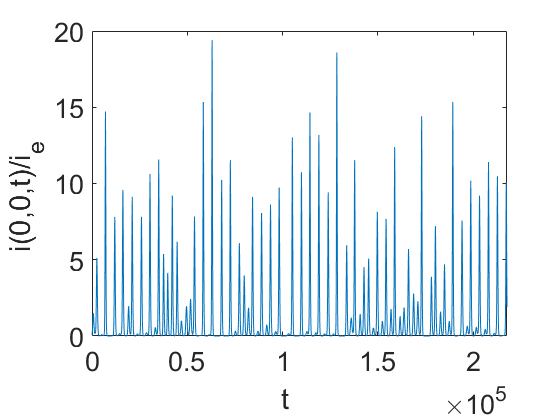}}
		\subfigure[]{\includegraphics[scale=0.35]{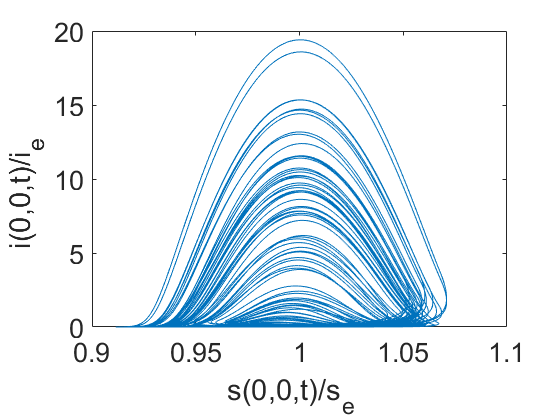}}}
		
			\mbox{\subfigure[]{\includegraphics[scale=0.35]{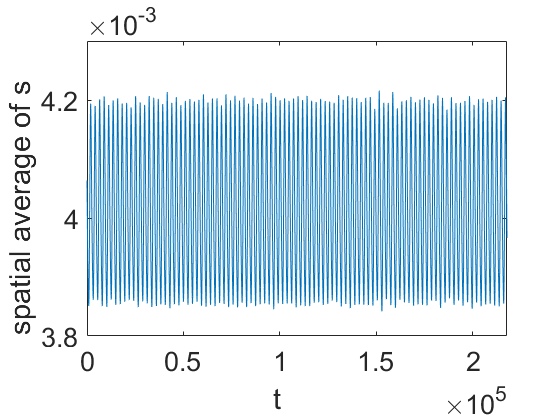}}
		\subfigure[]{\includegraphics[scale=0.35]{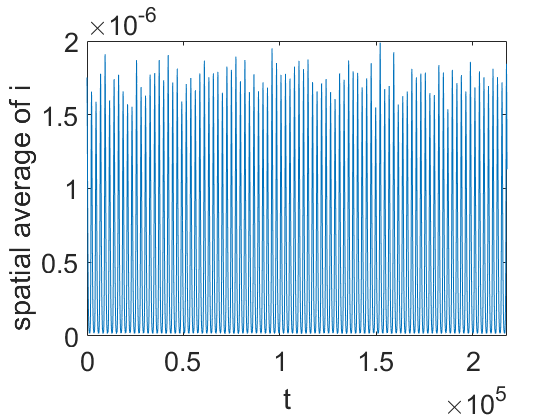}}
		\subfigure[]{\includegraphics[scale=0.35]{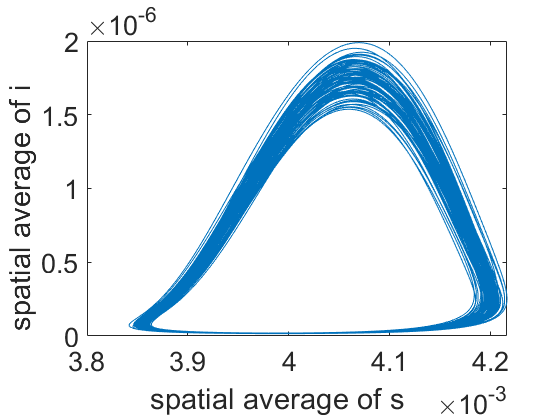}}}
\caption{Transition to spatio-temporal chaos. Impact of parameter $a$ on the dynamic behavior, in the bidimensional case. Here $a=0.01$. First and second rows show snapshots at three different times of the densities of, respectively, susceptible and infectious subjects, normalized to their respective endemic values. Third row shows time series and phase portraits computed at the spatial point $(0,0)$. Fourth row shows time series and phase of the averages state variables. Namely: panels (a), (d) correspond to snapshots of normalized value of $s(x,t)/s_e$  and $i(x,t)/i_e$ captured at time $t=250000$; panels (b), (e) show the same snapshots but  captured at time $t=252000$; finally panels (c), (f)  shows the snapshots captured at time  $t=254000$. Parameter values:\;
$L=1000$, $D=0.05$, $\mu =1/(75\times365)$, $\nu = 1/7$, $\beta=1.43$, $\rho = 8$, $p_0=0.5$, $s_e=0.1$, $i_e=0.000011$, $h=100$ and $a=0.01$. The initial perturbation is: $s(x,y,0)=s_e+0.01\xi_{xy}^1$, $i(x,y,0)=i_e+0.00001\xi_{xy}^2$  where $\xi_{xy}^j$, ($j=1,2$) are two spatially uncorrelated white noise terms. Periodic boundary conditions are used.}
\label{chaos2Da0p01}
\end{center}
\end{figure}

Now we describe the numerical simulation results for the model  (\ref{tpbspSefk})-(\ref{tpbspIefk})-(\ref{tpbspMefk}). We consider $p(M)=p_0 +\min\left(  c*M , 1-p_0 \right)$, $L=1000$, $D=0.05$, $\mu =1/(75\times365)$, $\nu = 1/7$, $\beta=1.43$, $\rho =8$, $p_0=0.5$, $h=100.$ \\
We assume as a key bifurcation value the delay-related parameter $a$, which assumes the following values:  $a=0.6,\; 0.1, \; 0.033, \; 0.01.$ 
Figures ~\ref{fig:exponentially_fading_1D_pattern} and ~\ref{New1D} show the impact of $a$ in the case of one dimension. Similarly, the impact of $a$ in the bidimensional case is illustrated by figures ~\ref{Turing2Da0p6} to Fig.~\ref{chaos2Da0p01}. In both cases, a transition from a Turing Pattern for $a=0.6$ to spatio-temporal chaos for lower values of $a$ is observed (see next section). Moreover, it is of interest to note that for comparatively small values of $a$  the patches with a large number of infected individuals increase: despite the fact that $i_e$ does not depend on $a$, However the maximum value of the ratio $i(x,t)/i_e$ increases gradually as $a$ decreases.  
%


\FloatBarrier


\section{Assessment of the chaotic nature of the simulated dynamics}\label{asse}

The change in dynamics of the spatial pattern from stationary Turing pattern to irregular 'chaos-like' time-varying patterns can be qualitatively understood from the figures~\ref{New1D}, ~\ref{chaos2Da0p1},\ref{chaos2Da0p033}, \ref{chaos2Da0p01}. However, this irregularity might not necessarily correspond to a chaotic nature of the dynamics and a detailed quantitative analysis is needed. To ensure the chaotic nature of these solutions, we apply techniques developed in the global study of spatio-temporal chaos developed in areas such as population dynamics \cite{malchow,morozov} and nonlinear physics \cite{pastur,vulpiani}. \textbf{In the appendix, we will also briefly apply  the statistical theory of nonlinear time-series analysis \cite{chaos_book_1,tserieschaos,volatile} to a local time-series computed at spatial point $(0,0)$.}\\
In our assessment we will refer to the parametric configuration and initial conditions that generated \textbf{the spatially bidimensional} simulations depicted in figure \ref{chaos2Da0p01} \textbf{and, for the 1D case, the lower panel of figure 8}.

\subsection{Estimating the Maximum Lyapunov Exponent}

\begin{figure}[thpb]
\begin{center}
		\mbox{\subfigure[]{\includegraphics[width=8cm,height=6cm]{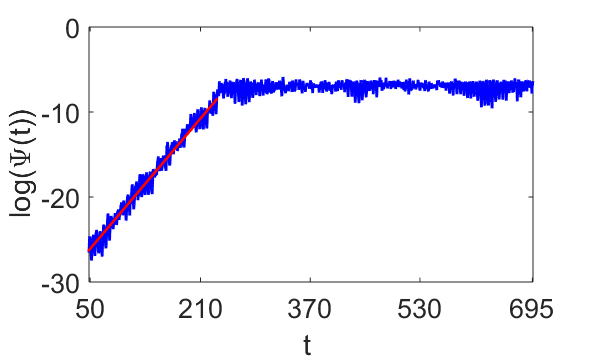}}
\subfigure[]{\includegraphics[width=8cm,height=6cm]{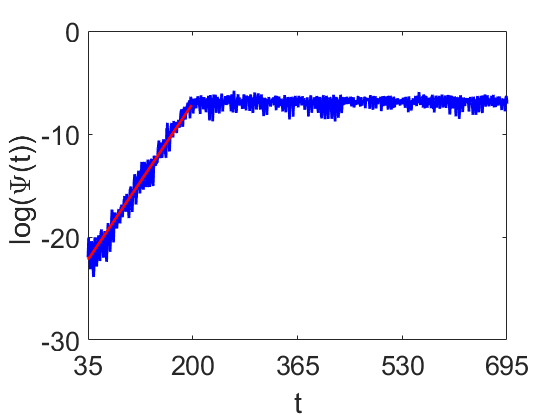}}
		}
\caption{Plot of $\log(\Psi(t))$ corresponding to Figure \ref{chaos2Da0p01} (in blue in Left panel) and third row of Figure \ref{New1D} (in blue in Right panel). The regression line  (in red) in both cases corresponds to the phase of growth. Description of the parameter values is given in the text.}
\label{population_dynamics_lyapunov_1}
\end{center}
\end{figure}

\textbf{The determination of the Maximum Lyapunov Exponent (MLE) for finite dimensional dynamical systems usually relies on the Benettin-Galgani-Giorgilli-Strelcyn algorithm \cite{benettin} which is, however, only applicable to non-dissipative systems \cite{pechuk2022maximum}. More empirical statistical physics-based numerical approaches are employed \cite{pechuk2022maximum,ndb}, especially in the context of spatiotemporal setting \cite{vulpiani,malchow,morozov}. In particular,} we start by following \cite{malchow,morozov} by  first considering the impact of \textbf{a single} small and localized perturbation on the dynamics of the system. Namely, we consider the following perturbed initial condition:
$$ s_{alt}(x,y,0)= s(x,y,0) $$
$$ i_{alt}(x,y,0)= i(x,y,0)\Big(1 + \varepsilon \sin\big(a x  \big)\cos\big(b y  \big) \Big)$$
with $\varepsilon \ll1$, and \textbf{$a=b=2\pi/L$ , where L=1000}.\\
We denote as $(s_{alt}(x,y,t), i_{alt}(x,y,t))$ the solution of the model corresponding to the above perturbed initial conditions.\\
We thus define the average normalized difference between the prevalences as follows:
$$ \Psi(t)= \frac{1}{i_e L^2}||i(x,y,t)-i_{alt}(x,y,t)||^{\Omega}_{2}. $$
where
$$ ||f(x,y,t)||^{\Omega}_{2} =\sqrt{ \int_{\Omega}f^2(x,y,t)\dif x \dif y} .$$
\textbf{In other words, $\Psi(t)$ is the functional distance in the space $\mathcal{L}^2$ between a reference prevalence density function $i(x,y,t)$ and its initially small perturbation $i_{alt}(x,y,t) $. We focus on the distance between the prevalence density functions because the epidemiological significance of the prevalence.} The function $\Psi(t)$ is such that if the system is chaotic then it has an exponentially divergent initial phase (typically followed by a plateau due to the boundedness of the state variable $i$) that allows to compute the so-called Maximum Lyapunov Exponent (MLE) \cite{ndb}.\\
Discarding the initial transients, the plot of $\log(\Psi(t))$ is shown in fig.~\ref{population_dynamics_lyapunov_1}(a) for time $t \in (50, \; 695)$, where the time is taken in years. We take $a=b=2\pi/L$ and $\epsilon=0.001$. \textbf{The estimated value of the MLE is} $\Lambda=0.0952 \; \textrm{years}^{-1}$ \textbf{with} $CI=(0.0879, 0.1025) \; \textrm{years}^{-1}$, \textbf{corresponding to a characteristic divergence time (defined simply as the inverse of the MLE) of} $10.5 \; \textrm{years}$, \textbf{with the following Confidence Interval} $CI=(9.761, 11.37) \; \textrm{years}$.
\textbf{Applying the above procedure to the 1D case, see panel b of the Figure \ref{population_dynamics_lyapunov_1}, we got the estimate for the MLE:} $\lambda=0.0922 \; \textrm{years}^{-1}$ \textbf{with} $CI=(0.0846, 0.0998) \; \textrm{years}^{-1}$,  \textbf{corresponding to a characteristic divergence time of} $10.8  \; \textrm{years}$\textbf{, with} $CI=(10.016, 11.818) \; \textrm{years}$.\\

\FloatBarrier

\textbf{In \cite{medvinsky2001patchy} Medvinsky and coauthors slightly generalized the above--described procedure by applying it four times from four distinct perturbed initial conditions. In this way, they  showed that \cite{medvinsky2001patchy}, for their model, the four estimated MLEs lied in a small range. Note that in \cite{medvinsky2001patchy} the Confidence Intervals for the four estimates of the MLE are not provided. Here, we propose another heuristic but more robust approach. Namely, we consider a sufficiently large number $N>>1$ of perturbed initial conditions, and for each of them we compute not only the estimated MLE but its CI. Finally, from the available $N$ estimates we derive an overall estimate for the MLE and for its CI. }\\  
\textbf{We indicate as}
$$ u(x,y,t) =\Big( s(x,y,;t), i(x,y,t), m(x,y,t) \Big)$$
\textbf{and a reference initial condition 
$u_{ref}(x,y,0)$ and the corresponding reference orbit $u_{ref}(x,y,t)$.}\\
\textbf{Furthermore, we consider $N$ small random perturbations of $u_{ref}(x,y,0)$: }
$$ u_{alt}^{(1)}(x,y,0),\dots,u_{alt}^{(N)}(x,y,0) $$
\textbf{where}
$$ s_{alt}^{(j)}(x,y,0)= s_{ref}(x,y,0) \big(1 + \epsilon_S^{(j)}(x,y)\big) $$
$$ i_{alt}^{(j)}(x,y,0)= i_{ref}(x,y,0)\big( 1 + \epsilon_I^{(j)}(x,y)\big) $$
$$ m_{alt}^{(j)}(x,y,0)= m_{ref}(x,y,0) =0.$$
\textbf{from the corresponding orbits }
$$ u_{alt}^{(1)}(x,y,t),\dots,u_{alt}^{(N)}(x,y,t) $$
\textbf{we compute the N $L^2$ functional distances }
$$ \eta^{(k)}(t)=||  u_{alt}^{(k)}(x,y,t)-u_{ref}(x,y,t) ||_2, \; k=1,\dots,N. $$
\textbf{Finally, from the 'linear zone' of each $\log (\eta^{(k)}(t))$ we can thus calculate and estimate of the MLE and of its standard error:}
$$ (\Lambda^{(1)},\sigma^{(1)}),\dots, (\Lambda^{(N)},\sigma^{(N)}) .$$
\textbf{This yields the following global estimate for the MLE and its associated standard error:}
$$ \Lambda^* = \frac{1}{N}\sum_{k=1}^{N}\Lambda^{(k)}, \;\;\;\; \sigma^* = \frac{1}{N}\sum_{k=1}^{N}\sigma^{(k)}.$$
\textbf{We start from the one spatial dimensions case, where we considered perturbations of the type}
$$ \epsilon_S^{(j)}(x) =0, \;\;\;\; \epsilon_I^{(j)}(x) = B \sum_{k=1}^{10} A_k^j \sin\big(k \frac{2\pi}{1000}x +\phi_k^j \big) ,$$
\textbf{where $B=0.001$ each $ A_k^j $ is a Random Variable uniformly distributed in (0,1); each $ \phi_k^j$ is also random Variable uniformly distributed in $(-\pi,+\pi)$.}\\
\textbf{By setting $N=30$ we obtained the following estimates:}
$$ \Lambda^* =  0.1073 \;  \textrm{\textit{years}}^{-1} \;\;\;\; \sigma^* = 0.0032 \;  \textrm{\textit{years}}^{-1}$$
\textbf{implying the following conventional confidence interval:} 
$$ CI = (0.101, 0.1135) \;  \textrm{\textit{years}}^{-1} $$
\textbf{The obtained average characteristic separation time is thus} $9.319 \;  \textrm{\textit{years}}$ 
\textbf{with} $ CI = (8.804, 9.89) \;  \textrm{\textit{years}} $. \textbf{For the case 2D, we proceeded similarly. Assuming the following initial conditions}
$$ \epsilon_S^{(j)}(x) =0, \;\;\;\; \epsilon_I^{(j)}(x) = B \sum_{k=1}^{10} A_k^j \sin\big(k \frac{2\pi}{1000}x +\phi_k^j \big) + C_k^j \cos\big(k \frac{2\pi}{1000}y +\eta_k^j \big) ,$$
\textbf{where $B=0.001$,   $ A_k^j $ and $ C_k^j $ are Random Variables uniformly distributed in (0,1); each $ \phi_k^j$ and $ \eta_k^j$ are random Variables uniformly distributed in $(-\pi,+\pi)$.}\\
\textbf{By setting $N=30$ we obtained the following estimate:}
$$ \Lambda^* =  0.1087 \;  \textrm{\textit{years}}^{-1} \;\;\;\; \sigma^* = 0.0034 \;  \textrm{\textit{years}}^{-1}$$
\textbf{implying the following conventional confidence interval:} 
$$ CI = (0.1020, 0.1154)) \;  \textrm{\textit{years}}^{-1}. $$
\textbf{The obtained average characteristic separation time is thus} $9.2 \;  \textrm{\textit{years}}$ 
\textbf{with} $ CI = (8.66, 9.8) \;  \textrm{\textit{years}} $. \\
\textbf{Summarizing, both in the one dimension and in two dimensions, the method proposed here provides a more accurate estimate the MLE and of its CI. }

\subsection{Spatial and temporal correlations}
The above computed \textbf{estimates of the} Maximum Liapunov Exponent allowed us to i) confirm the temporal chaotic nature of the system; ii) quantify the temporal features of the spatio-temporal chaos in our simulations. It remains to quantify their spatial features by determining the correlation length, that is the typical space scale $L_{corr}$ such that the time series of the system at two random point whose distance is bigger than $L_{corr}$ have no or very small correlation\cite{malchow,vulpiani}.

\begin{figure}[thpb]
\centerline{\includegraphics[width=11cm,height=6cm]{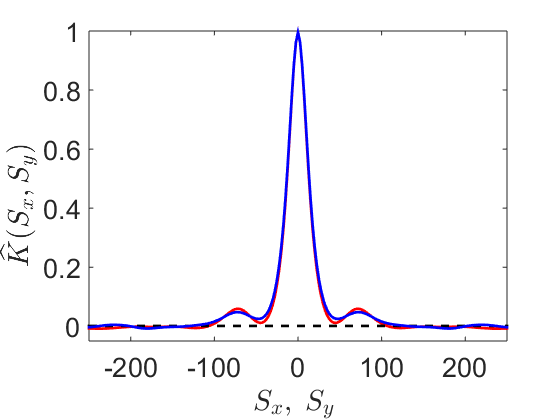}}
\caption{'Global Spatial Correlation Function' corresponding to Fig.~\ref{chaos2Da0p01} along the direction of axes X: $s=(s_x,0)$ (red curve) and along the direction of axes Y: $s=(0,s_y)$ (blue curve). $t_i= 20000$ days, $T_* = 254000$ days and all other parameters are as in Fig.~\ref{chaos2Da0p01}.}
\label{GSCFfig}
\end{figure}

In order to compute the correlation length we need to compute a measure of the spatial correlation \cite{pastur,malchow,morozov,vulpiani}. In literature there are many slightly different and 'correlated' definition of the spatial correlation function\cite{pastur,petrovskii2001wave,malchow,morozov,vulpiani}. 
We first adopt the definition used in \cite{pastur}, and then we will compare it with the definition used in \cite{morozov,petrovskii2001wave,malchow}. 
In both cases one has to compute the spatial average of the 'signal' (the disease prevalence, in our case):
$$ a(t)= \frac{1}{\meas(\Omega)}\int  i(x,y,t)\dif x \dif y;  $$
the difference between the signal $i(x,y,t)$ and its mean $a(t)$
$$ v(x,y,t) = i(x,y,t) - a(t), $$
and the variance, spatial or temporal (see later), of the signal.\\
The `Global Spatial Correlation'  function (GSCF) is given by \cite{pastur}
$$ \widetilde{K}(s_x,s_y)= \left\langle \frac{\big\langle v(r,t)v(r+s,t) \big\rangle_{r}}{ \big\langle v^2(r,t) \big\rangle_{r}}\right\rangle_{t} = \lim_{T \rightarrow +\infty} \frac{1}{T}\int_{t_i}^{t_i+T}\frac{N(s_x,s_y,t)}{\sigma^2(t)}\dif t $$
where $r=(x,y)$, $s=(s_x,s_y)$, $\langle \phi(r, t) \rangle_{\zeta} $ with $\zeta \in \{r,t\} $ denotes the average of the function $\phi(r, t)$ with respect to the variable $\zeta$, and 
$$  N(s,t)=  \frac{1}{\meas(\Omega)}\int v(r,t)v(r+s,t)\dif x \dif y $$
and $\sigma^2(t)$ is the spatial variance:
$$ \sigma^2(t)= \big\langle v^2(r,t) \big\rangle_{r}=\frac{1}{\meas(\Omega)}\int  v^2(r,t)\dif x \dif y$$
Theoretically, the GSCF  ought to be independent  of the direction of the vector $(s_x,s_y)$, i.e., it ought to be a function of $|s| =\sqrt{s_x^2+s_y^2}$. This is not always the case (see for an example \cite{pastur}). Thus, as in \cite{pastur} we plot the GSCF function along the two orthogonal directions corresponding to the axes: $(s_x,0) $ and $ (0,s_y)$. The auto-correlation lengths, along the directions $x$ and $y$ are approximated by the first zeros of the auto-correlation function along such directions.\\
The 'Global Spatial Correlation Function' corresponding to Figure \ref{chaos2Da0p01} is shown in Fig.~\ref{GSCFfig} along the direction of axes X: $s=(s_x,0)$ (blue curve) and along the direction of axes Y: $s=(0,s_y)$ (red curve). In practice the limit is approximated by computing the GSCF for a large but finite time $T_*$. Namely, we used $t_i=100000$ and $T_*=154000$. The correlation lengths are empirically identified with the first zeros of the GSCF, so that $L^{corr}_x \approx 131$ and $L^{corr}_x \approx 105$. Note however that the $K(0,s_y)$ reach a value very close to zero well before, namely around $S_y\approx 60$.

\begin{figure}[thpb]
\includegraphics[width=0.48\textwidth]{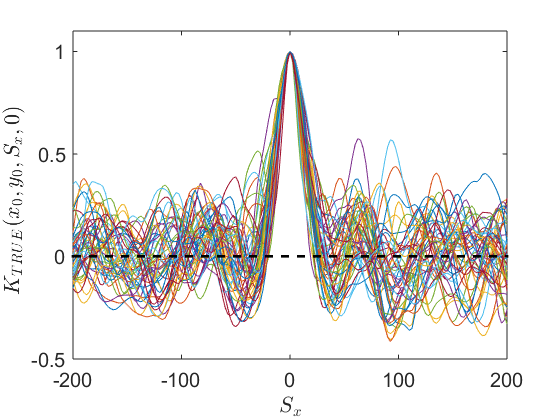}
\includegraphics[width=0.48\textwidth]{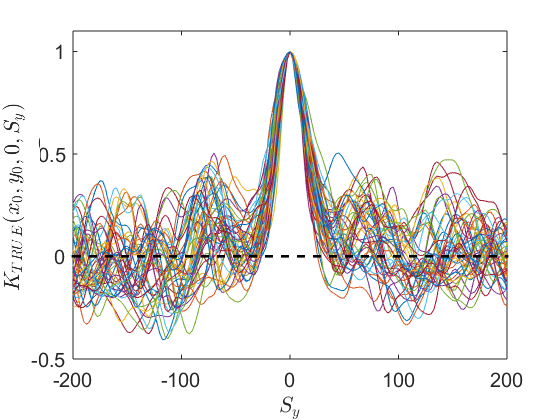}
\caption{50 instances of 'Two-points Spatial Correlation Function' corresponding to Fig.~\ref{chaos2Da0p01}, computed with 50 randomly chosen points $(x_0,y_0)$, along the direction of axes X: $s=(s_x,0)$ (left panel) and along the direction of axes Y: $s=(0,s_y)$ (right panel). $t_i= 20000$ days, $T_* = 254000$ days and all other parameters are as in Fig.~\ref{chaos2Da0p01}.}
\label{TSCFfig}
\end{figure}

A limitation of the GSCF is that it requires two averages: one spatial and one temporal. This makes the GSCF smooth but quite conservative in its estimate of $L_{corr}$. Another measure of spatial correlation widely used in statistical physics \cite{vulpiani} and in theoretical population biology \cite{petrovskii2001wave,malchow,morozov} is the so-called  'Two Points Spatial Correlation' function (TPSCF) that is defined as follows\cite{petrovskii2001wave,malchow,morozov,vulpiani}
$$
K_{TRUE}(r,s)=\frac{\Big\langle v(r,t)v(r+s,t) \Big\rangle_{t}}{\sqrt{\Big\langle v^2(r,t)\Big\rangle_{t}}\sqrt{\Big\langle v^2(r+s,t)\Big\rangle_{t}}}    
$$
where $r=(x_0,y_0)$
$$ \Big\langle f(x,y,t) \Big\rangle_{t}=lim_{T\rightarrow +\infty}\frac{1}{T}\int_{t_i}^{t_i+T}f(x,y,t)\dif t $$
So that
$$ K_{TRUE}(r,s)=lim_{T\rightarrow +\infty}\frac{ \frac{1}{T}\int_{t_i}^{t_i+T}v(r,t)v(r+s,t)\dif t}{  \sqrt{\frac{1}{T}\int_{t_i}^{t_i+T}v^2(r,t)\dif t} \sqrt{\frac{1}{T}\int_{t_i}^{t_i+T}v^2(r+s,t)\dif t} }.   $$
Also in this case, in the practice the above temporal limit is approximated as we did for the GSCF.\\
Theoretically, the spatial correlation ought to enjoy the following two properties:  i) to be independent  of the direction of the vector $(s_x,s_y)$ but only on $|s|=\sqrt{s_x^2+s_y^2}$, as  theoretically it ought to be for the GSCF; ii) to be independent of $r=(x_0,y_0)$. In the practice both properties frequently do not occur. \\
Operationally, we considered a set of randomly chosen values of $(x_0,y_0)$ and then computed and plotted the corresponding TPSCFs along the two orthogonal directions corresponding to the two axes, as shown in Figure \ref{TSCFfig}. We obtained the following results: 
\begin{itemize}
    \item Direction X positive values. Denoting as $Z^Y$ the first zero it is $median(Z^X)=30$, $mean(Z^X)=43.5$,  $sd(Z^x)=27.69$;
\item Direction X negative values. Denoting as $W^X$ the first zero: $median(W^X)=-31.25$, $mean(W^X)=-37.9$, $sd(W^X)=23.53$;
\item Direction Y positive values. Denoting as $Z^Y$ the first zero it is $median(Z^Y)=62.5$, $mean(Z^Y)=72.7$,  $sd(Z^Y)=50.5$;
\item Direction Y negative values. Denoting as $W^Y$ the first zero: $median(W^Y)=-50$, $mean(W^Y)=-61.78$, $sd(W^Y)=30.29$.
\end{itemize}
These results strongly suggest that the auto-correlation lengths are smaller than the one estimated by means of the GSCF.

\begin{figure}[thpb]
\centerline{\includegraphics[width=10cm,height=6cm]{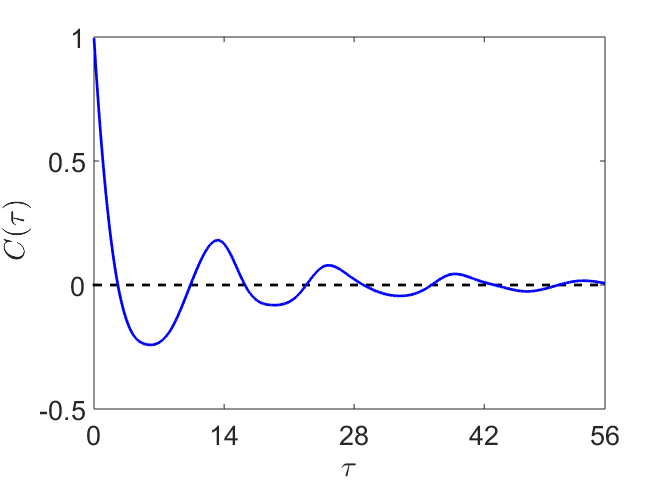}}
\caption{'Global Temporal Correlation Function' corresponding to Fig.~\ref{chaos2Da0p01}. $t_i= 20000$ days, $T_* = 254000$ days and all other parameters are as in Fig.~\ref{chaos2Da0p01}.}
\label{GTCFfig}
\end{figure}

Finally, in a spatio-temporal chaotic system the signal must also have low temporal correlation \cite{vulpiani}. This can be assessed by computing the 'Global Temporal Correlation' \cite{pastur}
$$ C(\tau)=\Big\langle \frac{\big\langle v(r,t)v(r,t+\tau) \big\rangle_{r}}{\sigma(t)\sigma(t+\tau)} \Big\rangle_{t} = \lim_{T \rightarrow +\infty}=\frac{1}{T}\int_{t_i}^{t_i+T} \frac{n(t,\tau)}{\sigma(t)\sigma(t+\tau)}\dif t $$
where
$$n(t,\tau)= \big\langle v(r,t)v(r,t+\tau) \big\rangle_{r}=\frac{1}{\meas(\Omega)}\int_{\Omega}  v(x,y,t) v(x,y,t+\tau)\dif x \dif y$$
The 'Global Temporal Correlation' in our case is shown in figure \ref{GTCFfig}. It shows a very rapid decay of the global temporal correlation. Namely, the halving time is approximately $\tau_{1/2} \approx 0.875$ years, whereas the temporal auto-correlation is null at time $\tau_0 \approx 2.573$ years.

\section{Concluding Remarks}
`Vaccine hesitancy' is a central topic in statistical physics of vaccination \cite{spva}. Here, we investigated, within a reaction-diffusion setting, a family of SIR models with vaccination and vaccine hesitancy, where agents' decisions depend on the available information a on the disease prevalence.
In particular, the core of our work has been the modelling of the spatio--temporal structure of the information used by agents to inform their  immunization--related decisions. This information is typically non-local over both its spatial and temporal components. \\
As main theoretical results we showed that the use of non--local information can generate a rich dynamics ranging from Turing patterns to Turing-Hopf bifurcations to spatio--temporal chaos. A remarkable exception leading to stability is, instead, represented by the case where the information used by the agents is global: in such a case we showed that the spatially homogeneous endemic equilibrium remains locally stable.
Extensive numerical simulations were carried out to validate the theoretical results and to deepen the analysis of patterning. 

In particular, We found that the spatial distribution of susceptible and infected individuals changes significantly
with the (average) duration of the memory involved with the temporal kernel. Short memories induce stationary pattern generated through Turing instability, whereas long-lasting memories lead to time variability in the spatial distribution of susceptible and infected subjects and to spatio-temporal chaos.\\

Note that \textbf{although} on the one hand  our spatial SIR model has a nonlinear multiplicative term $\beta s i$ due to mass action law typical of Lorenz model and of so many chaotic models in population biology \cite{malchow,murray2002intro,murray2002spatial}, on the other hand here the route to chaos is deeply different from the one observed in non-spatial chaotic SIR models. Indeed, in that models the transmission rate is periodic, whereas in \textbf{the model investigated here} $\beta$ is constant. \textbf{Here the onset of} the spatio-temporal chaos is linked to vaccine hesitancy \textbf{and not to the periodic variation of the contact rate. Namely, the chaos arises due to the interplay between the spatial diffusion and both the temporal non-locality and the spatial non-locality in the model of the spatio-temporal information index. Finally, we remind the readers that the non-spatial SIR model with vaccine hesitancy \cite{domasa} chaos does not onset even in the presence of the temporal non-locality in the information index, see} \cite{domasa}.


 At the best of our knowledge, this is the first time where it is shown that the introduction of vaccine hesitancy may induce spatio-temporal chaos and other simpler patterns in an epidemic model.\\
 From the Public Health (PH) viewpoint, this result is of interest and in line with the increasing relevance of spatial and spatio-temporal statistics \cite{kent2022spatial,lawson2018bayesian} in epidemiology of epidemic and endemic infectious diseases \cite{kramer2010modern,abubakar2016infectious,meliker2011spatio,sun2016pattern,chen2014analyzing,lawson2018bayesian,haining2020regression,tang2017spatial,yu2018spatial}. Specifically, a spatially homogeneous endemic equilibrium that is stable means an epidemiological state 'simple' to manage because in the whole area that PH authorities must monitor the disease is uniformly distributed. A static pattern means that the disease is distributed in a non-homogeneous way with local peak that potentially cannot be managed due to limited resources and to logistic difficulties \cite{kramer2010modern,abubakar2016infectious,meliker2011spatio,sun2016pattern,chen2014analyzing,tang2017spatial,yu2018spatial}. Moreover, patterns contribute to geographical inequalities \cite{tang2017spatial,yu2018spatial}: specific and expensive measures targeting the high-risk areas corresponding to clusters have to be developed \cite{yu2018spatial}. As a consequence, spatio-temporal chaos is the most complex scenario for PH since it means that clusters move, appear, and disappear in a pseudo--random fashion. This may make the management of an endemic scenario extremely complex since in the practice there are continuous recurrent epidemics that are spatially uncorrelated and temporally irregular. This is a great obstacle to the planning of allocation of resources. \\
 
 
 Another point is of interest for PH: we showed that there is a transition between spatial patterns to spatio-temporal chaos, which is determined by the average information delay. The study of transitions between spatio-temporal behaviors is important for PH to the aim of a control of early warning signs of imminent outbreak \cite{sun2016pattern}. This of course will require a further research step.\\
 
 Finally, from the viewpoint of statistics providing evidences of clustering and spatio-temporal chaos both induced by behavioral mechanisms could suggest new kinds of analyses for existing and future epidemiological data.

\textbf{From a computational viewpoint, we have introduced here a very simple heuristic algorithm to estimate the MLE, and its confidence interval, of a given Dynamical System. The properties of the algorithm will be investigated elsewhere with reference to other algorithms, as the one proposed in \cite{benettin,pechuk2022maximum,ndb}, and in the context of other specific models.}

\textbf{From a more mathematical viewpoint, two problems are of particular interest, in our opinion. On the one hand, here we have only numerically investigated the onset of spatio-temporal chaos in the model we have proposed. Thus, an analytical study would be welcome. On the other hand, here we have stressed an important cause triggering spatio-temporal chaos in the behavioral SIR model. Many related problems remain open. For example, as implicitly suggested by one of the referees, it would be important to investigate the spatio-temporal dynamic effects of a seasonally varying contact rate in the context of the spatio-temporal SIR model } 

The reaction-diffusion framework we have adopted suffers some drawbacks in view of its simplistic representation of human mobility. This is a clear limitation of our study. Nonetheless, this approach has some non-trivial advantages compared to other types of models. The first one is that when we investigate fundamental conceptual problems, such as the stability of spatially homogeneous endemic states, a simple spatial population dynamics allow to tackle fully meaningful questions in a simple way, often analytical. Moreover, it is easy to show that the use of more realistic hyper-diffusive model of spatial mobility would any case lead to analytical Local Stability conditions that are a trivial extension of the one we have investigated here. Finally, another limitation of our investigation is that we performed all the simulations by assuming periodic boundary conditions to avoid boundary effects caused by the nonlocal kernel. We will investigate the general case in the next future. 

\section*{Acknowledgments}
We warmly thank the two anonymous referees and the Associated Editor for their very important suggestions that helped us to substantially improve this work. The work by AdO has been done under the auspices of the Gruppo Nazionale di Fisica Matematica (GNFM) of the Italian Istituto Nazionale di Alta Matematica (INDAM).

\bigskip

\appendix
\noindent
\textbf{\Large{APPENDIX}}
\section{Temporal and Spatio-temporal SIR model with constant vaccine uptake}
Here we briefly summarize the standard SIR model for childhood mandatory immunization at birth, which reads as follows
\begin{align}
\frac{d}{d t}S&=\mu N(1-p) -\mu S - \beta(t)  \frac{I}{N} S, \nonumber\\
\frac{d}{d t}I&=\beta(t) \frac{I}{N} S - (\nu+\mu) I, \nonumber\\
\frac{d}{d t}R&=\nu I -\mu R, \nonumber\\
\frac{d}{d t}V&=\mu N p - \mu V, \nonumber
\end{align}
where $S(t)$, $I(t)$, $R(t)$ denote the number of susceptible,  infectious, removed and vaccinated subjects, respectively; $N(t)=S(t)+I(t)+R(t)+V(t)$ denotes the total population size, $\mu$ represents both the mortality rate and the birth rate, therefore ensuring that the population size is stationary over time i.e., $N(t)=N$ at all times, $\beta(t)$ is the transmission rate of the disease, which we will assume either constant or periodically varying with period one year (due to yearly recurrent social phenomena, as the school calendar, and/or to weather seasonality)\cite{bbnnffReview}, and $0<p<1$ the effective immunization uptake at birth, taken as a constant (i.e., behavior-free) in the standard model. \\
Letting 
$$\overline{\beta}= \frac{1}{T}\int_0^T\beta(x)\dif x$$ 
if it is:
\begin{equation}\label{MayAndrsonCondition}
 (1-p)\frac{\overline{\beta}}{\mu+\nu} <1 \Rightarrow	p> p_{cr} = 1-\frac{1}{BRN},
\end{equation}
(where $BRN=\overline{\beta}/(\mu+\nu)$ denotes the SIR \emph{basic reproduction number}) then the disease-free equilibrium (DFE)
$$ DFE = (N(1-p),0,0,p N)$$
is globally asymptotically stable (GAS).\\

Let us now include the impact of spatial heterogeneity by plugging the previous standard model for childhood immunization within the most basic model for human mobility with explicit space, namely the standard PDE diffusion model. Let now $S(x,t),I(x,t),R(x,t),V(x,t)$ denote the state variables representing the absolute spatial densities of susceptible, infective, removed, vaccinated occupying position $x$ at time $t$, respectively, with 

$$S(x,t)+I(x,t)+R(x,t)+V(x,t)=N(x,t).$$

The corresponding model is described by the following system of partial differential equations (PDEs)
\begin{align}
\partial_t S&=D \nabla^2 S +\mu N(1-p) -\mu S - \beta(t)  \frac{I}{N} S, \nonumber\\
\partial_t I&=D \nabla^2 I +\beta(t) \frac{I}{N} S - (\nu+\mu) I ,\nonumber\\
\partial_t R&=D \nabla^2 R +\nu I -\mu R ,\nonumber\\
\partial_t V&=D \nabla^2 V +\mu N p - \mu V ,\nonumber\\
\partial_t N &= D \nabla^2 N ,\nonumber
\end{align}
to be solved in a bounded spatial set $\Omega$ under Neumann-type boundary conditions
$$ \partial_n S =\partial_n I =0 \textrm{  } x \in \partial \Omega .$$
As in the non--spatial model, we assume that the population is at equilibrium
$$ \lim_{t\rightarrow +\infty}n(x,t)= Z = <n(x,0)>. $$
Defining the location-specific epidemiological fractions $s(x,t),i(x,t),r(x,t),v(x,t)$ such that
$$ (S,I,R,V)  =Z(s,i,r,v) , $$
we get,
\begin{align}
\partial_t s&=D \nabla^2 s +\mu (1-p) -\mu s - \beta(t)  i s, \nonumber\\
\partial_t i&=D \nabla^2 i +\beta(t) i s - (\nu+\mu) i, \nonumber
\end{align}
We have omitted the equations for $r$, which is linear, and $v$, since $v=1-s-i-r$.\\
From
$$ \partial_t s \le D \nabla^2 s +\mu (1-p) -\mu s  $$
it follows that asymptotically
$$ s(x,t) \le 1 -p.$$
Thus,
$$ \partial_t i \le D \nabla^2 i +\left( (1 -p)\beta(t)  - (\nu+\mu)\right) i, $$
implying that if (\ref{MayAndrsonCondition}) holds then
$$ \lim_{t \rightarrow \infty}(s(x,t),i(x,t))=(1-p,0).  $$
The latter result proves that the condition (\ref{MayAndrsonCondition}) continues to ensure that the $DFE$ remains GAS even when the model is spatially structured. If
\begin{equation}\label{unbstMayAndrsonCondition}
 (1-p)\frac{\overline{\beta}}{\mu+\nu} >1 \Rightarrow	0<p< p_{cr} = 1-\frac{1}{BRN},
\end{equation}
the DFE is unstable, as it is easy to show by considering the linearized equation for the infectious:
$$ \partial_t w = D \nabla^2 w +\left( (1 -p)\beta(t)  - (\nu+\mu)\right) w. $$
Finally, if the transmission rate is  constant $ \beta(t) = \beta^*$ by setting
$$ (S,I)=(1-p)( \widehat{S},\widehat{I} )  $$
one gets that the dynamics of $( \widehat{S}, \widehat{I} ) $ is ruled by the following 'spatial SIR model without vaccination':
\begin{align}
\partial_t \widehat{S}&=D \nabla^2 \widehat{S} +\mu (1-\widehat{S})  - \beta^*  \widehat{I} \widehat{S}, \label{eqU}\\
\partial_t \widehat{I}&=D \nabla^2 \widehat{I} +\beta^* \widehat{I} \widehat{S} - (\nu+\mu) \widehat{I}, \label{eqW}.
\end{align}
Proceeding as in \cite{thSIR}, it is easy matter to show that if (\ref{unbstMayAndrsonCondition}) holds then system (\ref{eqU})-(\ref{eqW}) has a unique GAS constant uniform endemic equilibrium $(\widehat{S}_e,\widehat{I}_e)$. As a consequence also the original model has a unique GAS constant uniform endemic equilibrium 
$$(1-p)(\widehat{S}_e,\widehat{I}_e)=\Big(\frac{1}{BRN},\frac{\mu}{\mu+\nu}(p_{cr}-p)\Big)$$.
\section{Fourier Transform definition}
In this work we adopt the following definition of Fourier transform of a suitable function $f(y)$:
$$
\widetilde{f}(\xi) = \int_{\mathbb{R}^n}f(y)e^{-i\xi y}\dif x.
$$
\section{Statistical analysis of time series corresponding specific points in the domain $\Omega$}
\textbf{In the section \ref{asse} the assessment of the MLE and of the  measures of spatial and temporal correlation confirmed the chaotic nature of our simulations.}\\
However, as one can see in Figures~\ref{New1D}, ~\ref{chaos2Da0p1},\ref{chaos2Da0p033}, \ref{chaos2Da0p01} the time-series computed at point $(0,0)$ are, qualitatively speaking, 'more irregular' than those of the average values\textbf{, which are also shown in the same figures}. Thus it is of interest to briefly analyze these local time series as if they were time series of temporal signals of unknown origin. We made this analysis by using the R library \textit{tseriesChaos} \cite{tserieschaos,volatile,chaos_book_1}.

\begin{figure}[thpb]
\begin{center}
		\mbox{\subfigure[]{\includegraphics[scale=0.4]{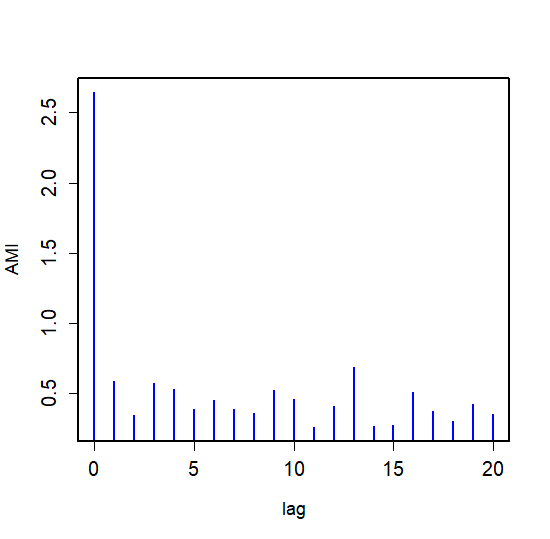}}.
		\subfigure[]{\includegraphics[scale=0.4]{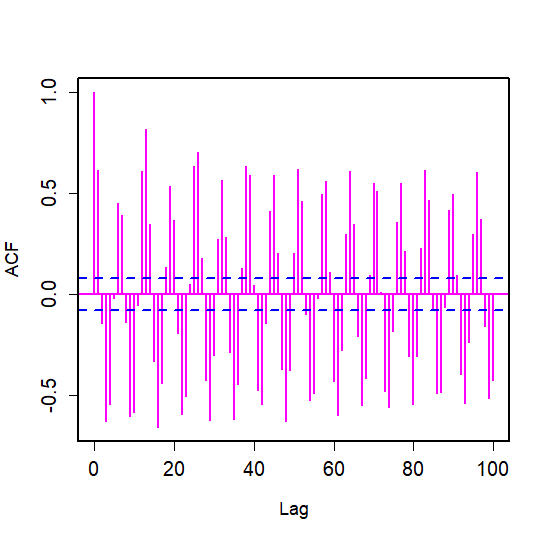}}
		}\\
		\mbox{\subfigure[]{\includegraphics[scale=0.4]{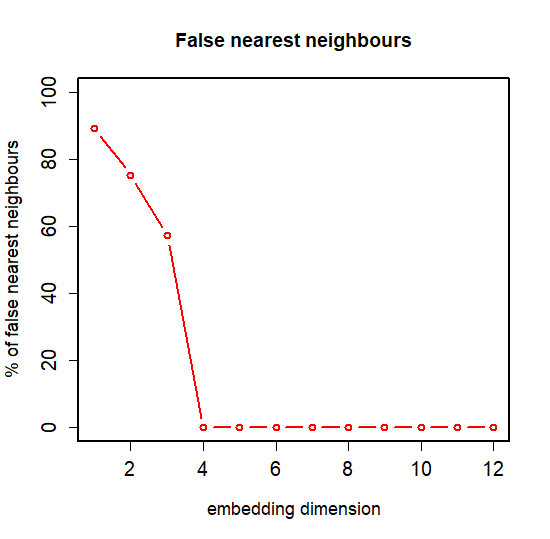}}
		\subfigure[]{\includegraphics[scale=0.4]{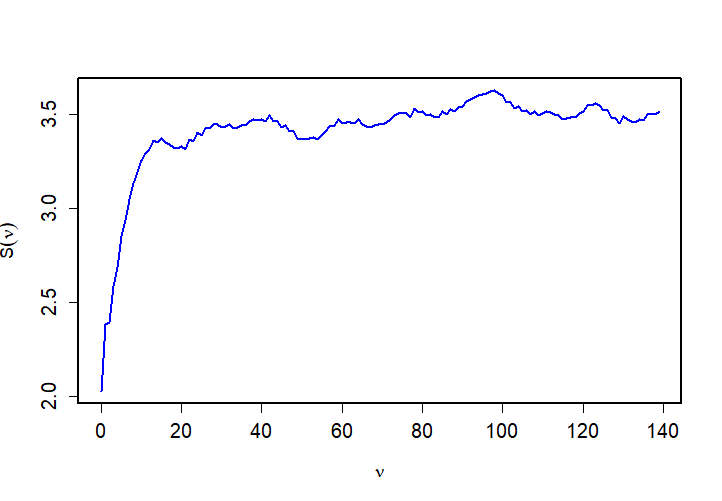}}
		}
\caption{Statistical assessment of the chaotic nature of the normalized time series $S_N$ (a) AMI function, suggesting an embedding delay $d=2$; (b) Auto-correlation function, suggesting a Theiler window $tw=3$; (c) Percentage of false nearest neighbours, suggesting an  embedding dimension $m=4$. (d) Lyapunov diagram with $(d, m, tw)=(2, 4, 3).$  }
\label{chts}
\end{center}
\end{figure}

We considered the re-scaled time-series data for the susceptible subjects at $(0,0)$ $\hat{s} = s(0, 0, t)/s_e$ shown in Fig.~\ref{chaos2Da0p01}, which we normalized:
$S_N=\hat{s}-\mu_{\hat{s}}/\sigma_{\hat{s}},$ where $\mu_{\hat{s}}=Mean(\hat{s})$ and $\sigma_{\hat{s}}=sd(\hat{s})$.\\
The average mutual information plot \cite{chaos_book_1,chaos_analyis_1,giannerini2012quest,giannerini2004assessing} of $S_N$ (upper left panel of Fig.~\ref{chts}) has its first relative minimum at $2$, which estimates the embedding delay \cite{chaos_book_1,chaos_analyis_1} $d=2.$ Then, the first minima of the auto-correlation function (upper right panel of Fig.~\ref{chts}) indicates the following estimate of the Theiler window \cite{chaos_book_1,chaos_analyis_1}: $tw=3$. The false nearest neighbour plot (lower left panel of Fig.~\ref{chts}) has its first minima at $4$, corresponding to the following estimated embedding dimension \cite{chaos_book_1,chaos_analyis_1} $m=4$. These three estimates allow to plot the Lyapunov diagram corresponding to the time-series $S_N$ (lower right panel of Fig.~\ref{chts}) and to estimate, from its linear part, the Maximum Lyapunov Exponent (MLE) $\Lambda \approx 0.109808 \; \textrm{years}^{-1}$ \textbf{is positive and its associate confidence interval reads as follows:} $CI =(0.09892804, 0.12068796) \; \textrm{years}^{-1}$   which confirms that time-series $S_N$ is chaotic. \textbf{This average MLE agrees with the one we have estimate by using the model.}\\
\FloatBarrier



le{plain}

\end{document}